\documentclass[paper]{JHEP3}
\JHEPspecialurl{http://jhep.sissa.it/JOURNAL/JHEP3.tar.gz}
\usepackage{epsfig,multicol}
\usepackage{cite}
\usepackage{amsmath,amssymb}
\usepackage{bm}
\usepackage{type1cm}

\preprint{KEK-TH-1145}
\title{
Exact fuzzy sphere thermodynamics in matrix quantum mechanics
      } 
\author{ 
Naoyuki Kawahara${}^{ab}$\,,
Jun Nishimura${}^{ac}$ 
and Shingo Takeuchi${}^{c}$ 
\vspace*{0.5cm} \\  
\llap{$^a$}High Energy Accelerator Research Organization (KEK),\\
Tsukuba, Ibaraki, 305-0801, Japan  \\
\llap{$^b$}Department of Physics, Kyushu University, %\\
Fukuoka 812-8581, Japan \\ 
\llap{$^c$}Department of Particle and Nuclear Physics,\\
Graduate University for Advanced Studies (SOKENDAI),\\
Tsukuba, Ibaraki, 305-0801, Japan
\vspace*{0.5cm} \\ 
\email{kawahara@post.kek.jp,
jnishi@post.kek.jp, shingo@post.kek.jp}
}

\abstract{
We study thermodynamical properties of a fuzzy sphere
in matrix quantum mechanics of the BFSS type
including the Chern-Simons term.
Various quantities are calculated to all orders in
perturbation theory exploiting the one-loop saturation
of the effective action in the large-$N$ limit.
The fuzzy sphere becomes unstable at sufficiently
strong coupling, and the critical point is obtained
explicitly as a function of the temperature.
The whole phase diagram 
is investigated by Monte Carlo simulation. 
Above the critical point, we obtain perfect agreement
with the all order results.
In the region below the critical point,
which is not accessible by perturbation theory,
we observe the Hagedorn transition.
In the high temperature limit our model is equivalent to 
a totally reduced model, and the relationship to
previously known results is clarified.
}
\keywords{Non-Commutative Geometry, Matrix Models,
Thermal Field Theory}

\def\be{\begin{equation}}
\def\ee{\end{equation}}
\def\beq{\begin{equation}}
\def\eeq{\end{equation}}
\def\bear{\begin{eqnarray}}
\def\eear{\end{eqnarray}}
\def\beqa{\begin{eqnarray}}
\def\eeqa{\end{eqnarray}}
\def\nn{\nonumber}
\def\del{\partial}

\newcommand{\dt}{\partial_t}

\newcommand{\ijk}{\epsilon_{ijk}}

\newcommand{\tk}{\tilde{\kappa}}
\newcommand{\ta}{\tilde{\alpha}}
\newcommand{\tb}{\tilde{\beta}}
\newcommand{\tT}{\tilde{T}}
\newcommand {\tr}{{\rm tr\,}}
\begin{document}
\section{Introduction}

Fuzzy sphere \cite{Madore}, 
which is a simple compact noncommutative manifolds,
has been discussed extensively in the literature.
One of the motivations comes from the general expectation 
that noncommutative geometry provides
a crucial link to string theory and quantum gravity.
Indeed Yang-Mills theories on noncommutative geometry
appear in a certain low energy limit of string theory \cite{Seiberg:1999vs}.
There is also an independent observation that the space-time 
uncertainty relation, which is naturally realized by noncommutative
geometry, can be derived from some general assumptions
on the underlying theory of quantum gravity \cite{gravity}.
As another motivation, fuzzy manifolds may be used
as a novel regularization method in quantum 
field theories \cite{Grosse:1995ar}.
%which may have certain advantage to the lattice
%
%% Unlike the lattice, fuzzy spheres preserve the continuous symmetries 
%% of the space-time considered, and hence it is expected that 
%% the situation concerning chiral symmetry 
%% and supersymmetry might be improved.
%
% in lattice field theories may become easier to overcome.
% In both contexts, it is important to study four-dimensional
% fuzzy manifolds eventually in order to 
% Although the simplest fuzzy sphere is the fuzzy 2-sphere,

In string theory, fuzzy spheres appear
as D-branes in the presence of 
external fields \cite{Myers:1999ps}.
In particular they appear as classical 
solutions\footnote{More general 
classical solutions such as a rotating
fuzzy sphere are discussed
in refs.\ \cite{Bak,HS1,Park,sol,HSKY}.}
in the pp-wave matrix model \cite{BMN}, which is
a generalization of the BFSS matrix theory \cite{9610043}
to the so-called pp-wave background \cite{KG}. 
Fundamental issues such as
the stability of these solutions \cite{DJR,SY3,HSKY}
and the spectrum of the fluctuations
\cite{DJR,DJR2,KP,KP2} have been 
studied intensively.\footnote{Similarly, fuzzy spheres
appear as classical solutions in
matrix string theory \cite{9703030}
on a type IIA plane-wave background \cite{SY4}.
The spectrum around the fuzzy spheres is computed in ref.\ \cite{DMS}.
%using the method of \cite{DJR}. 
This theory is used to study the matrix big bang \cite{DM}.}

Thermodynamical properties of the pp-wave matrix model 
have also been studied by various authors.
In ref.\ \cite{FSS} the free energy around the trivial vacuum,
which corresponds to a transverse M5-brane \cite{TM5} at
zero temperature\footnote{A fuzzy five-sphere
solution was constructed \cite{YL}
in a deformed plane-wave matrix model with an
interaction term due to the 6-form potential.},
was evaluated at the one-loop level, and the Hagedorn transition
%in the large-$N$ limit, where $N$ is the matrix size,
was studied in detail.
(See refs.\ \cite{SRV,Semenoff2} for a two-loop extension
and ref.\ \cite{Semenoff} for a review on this subject.) 
This calculation have been extended to more general vacua
in refs.\ \cite{Huang, HSKY-thermal,KNY}.
In all these works, however, a mass parameter, 
which parametrizes the deviation from the flat background,
is assumed to be large so that higher loop effects can be neglected.

In this paper we show that in fact 
it is possible to study the fuzzy sphere thermodynamics
to all orders in perturbation theory.
%all order calculation is possible around the fuzzy sphere
%configurations.
%
%backgrounds.
%certain cases.
While the method can be applied to more general models
including the pp-wave matrix model,
here we demonstrate it in a simple model, 
which can be obtained by dimensionally reducing 
a 4d U($N$) gauge theory to 1d.
The model has been used recently to compute
the mass gap in the theory of bosonic membranes \cite{Agarwal:2006sv}.
The action contains the 3d Chern-Simons term
representing the coupling to a background flux \cite{Myers:1999ps},
%which corresponds to the Myers term 
which enables fuzzy spheres to appear as classical solutions.
When the Chern-Simons coupling is switched off,
it
%the model 
reduces to the ``4d
bosonic BFSS matrix theory'' \cite{latticeBFSS}.
%as a particular case.
%
%studied earlier in refs.\ 
%where the coefficient of the Chern-Simons term ($\alpha$) is 
%set to zero.
Studying matrix quantum mechanics at finite temperature\footnote{More
generally, large-$N$ gauge theory at finite temperature
has been an active field of research \cite{Aharony,Aharony:2004ig,Aharony3,%
Alvarez-Gaume:2006jg,Lucini:2005vg}
partly motivated from the gauge/gravity correspondence.}
is itself
an interesting subject \cite{AMS,others,KLL,Aharony:2004ig},
in particular, because of its relation to the black hole physics 
\cite{KS,BFKS,Itzhaki:1998dd}.
%Such studies have been performed in refs.\ 
%The gaussian approximation advocated in ref.\ \cite{Kabat:2000hp}
%revealed interesting black hole thermodynamics \cite{blackholes}.

The method for the 
all order calculation has been developed
in totally reduced models \cite{0403242,ANN},
which are motivated in the context of 
the type IIB matrix model \cite{9612115}.
The key observation is that, in the large-$N$ limit,
the effective action is saturated 
at one loop in the bosonic case \cite{ANN}, and 
at two loop in the supersymmetric case \cite{0403242}.
Through the Legendre transformation, 
one can obtain the free energy and various observables
to all orders.
(We emphasize that this is different
from a typical situation in supersymmetric field theories,
in which higher loop corrections simply vanish due to cancellation.)
In the bosonic case, it has been confirmed that
the all order results are in perfect agreement
with the Monte Carlo results 
\cite{0401038} obtained in the fuzzy sphere phase.
These works have also been extended to
four-dimensional fuzzy manifolds \cite{0405277,0506205}.

Similarly to the results in ref.\ \cite{0401038},
we find in the present finite-temperature system that 
a single fuzzy sphere
becomes unstable at some critical $\alpha$, the coefficient 
of the Chern-Simons term.
This phenomenon occurs at any temperature,
and we obtain explicitly the critical $\alpha$ as a function of
the temperature.
%It turns out that in order to stabilize 
%the $T$ dependence of the critical $\alpha$ in the large-$N$ limit,
%we have to rescaled $\alpha$ and $T$ with appropriate powers of $N$.
We also perform Monte Carlo simulation
and confirm that the all order results for various observables
agree very well with the Monte Carlo results
above the critical $\alpha$.
% with the Monte Carlo simulation.
In the region below the critical $\alpha$, which is not
accessible by perturbation theory, we observe the Hagedorn transition
at some critical temperature.
At high temperature
our model is equivalent
to a totally reduced model, which is analogous to the model
studied in ref.\ \cite{0401038}.
%comparing the results at high temperature to those
%of the 0d matrix model,
%performing Monte Carlo simulation 
%We confirm this equivalence explicitly by Monte Carlo simulation.
We clarify the relationship to the results obtained there.

The rest of this paper is organized as follows.
In section \ref{sec:model} we define our model and
discuss its classical solutions.
%In section \ref{sec:boundary} we study...
In section \ref{section:pert-calc}
we show how one can perform the all order calculation
in perturbation theory.
In section \ref{sec:boundary}
we compare the all order results with the Monte
Carlo results.
In section \ref{sec:YM-Hage} we study the 
region in the phase diagram below the critical $\alpha$,
and show that the Hagedorn transition takes place.
In section \ref{sec:dim-red} we discuss
the high temperature limit of the model.
%In particular, we clarify the connection to the 
%known results in the totally reduced model.
Section \ref{sec:summary} is devoted to a summary
and discussions.

\section{The model and its classical solutions}
\label{sec:model}

The model we study in this paper is
defined by the action\footnote{We could have replaced 
the overall factor of $N$ in the action (\ref{action})
by $\frac{1}{g^2}$, where $g$ represents the Yang-Mills 
coupling constant. Our choice would then correspond to
setting the 't Hooft coupling $\lambda = g^2 N$ to unity.
We do not lose any generality, however, since the model
for arbitrary $\lambda$ can be readily obtained by
rescaling $X_i \rightarrow \lambda^{-1/3} X_i$,
$\beta \rightarrow \lambda^{1/3} \beta$,
$\alpha \rightarrow \lambda^{-1/3} \alpha$.
\label{footnote:tHooft}
} 
\beq
S = N \int_0^\beta \!\!dt \, {\rm tr} 
\left\{ \frac{1}{2} 
\Bigl( D_t X_i(t) \Bigr) ^2 -\frac{1}{4}
\Bigl( [X_i(t),X_j(t)] \Bigr)^2 
+\frac{2}{3} \, i \, \alpha \, \epsilon_{ijk}\, 
X_i(t) X_j(t) X_k(t) \right\} \ ,
\label{action}
\eeq
where $D_t$ represents the covariant derivative
%\beq
$
D_t =
\partial_t
- i \, [A(t),\hspace{0.2cm}\cdot\hspace{0.2cm}] \ .
$
%\nn 
%\eeq
The dynamical variables $A(t)$ and
$X_i(t)$ $(i=1,2,3)$ 
are $N \times N$ Hermitian matrices, which
can be regarded as the gauge field and 
three adjoint scalars, respectively, in a 1d gauge theory
with the U($N$) gauge symmetry
\beq
X_i(t) \rightarrow  g(t) \, X_i(t) \, g(t)^\dagger  \ ;
\quad 
A(t) \rightarrow  g(t) \, A(t) \, g(t)^\dagger 
+ i \, g(t) \, \frac{d}{dt}  \, g(t)^\dagger \ .
\label{gauge-sym}
\eeq
%% \beqa
%% X_i(t) &\rightarrow&  g(t) \, X_i(t) \, g(t)^\dagger  \ ,
%% \nn \\
%% A(t) &\rightarrow&  g(t) \, A(t) \, g(t)^\dagger 
%% + i \, g(t) \, \frac{d}{dt}  \, g(t)^\dagger \ .
%% \label{gauge-sym}
%% \eeqa
%Since the trace part of the gauge field decouples,
%we restrict $A(t)$ to be traceless in what follows.
The Euclidean time $t$ in (\ref{action})
has a finite extent $\beta$, 
which is related to the temperature $T$ through
$\beta = 1/T$,
%corresponds to the inverse temperature 
%$\beta\equiv 1/T$,
and all the fields obey periodic boundary conditions.
The cubic term represents the Chern-Simons term,
which is crucial for fuzzy spheres to become
classical solutions. The $\alpha = 0$ case
corresponds to the ``4d bosonic BFSS model''
studied in refs.\ \cite{latticeBFSS}.

The classical equations of motion 
can be obtained from the action (\ref{action}) as
%, which can be obtained
%by extremizing the action (\ref{action}), reads 
\beqa
%\label{eomX} 
 (D_t)^2 X_i &=& 
[X_j,[X_j,X_i]]
+ i \, \alpha \, \epsilon_{ijk}
[X_j,X_k]  \ , \\
%i[X_i,\partial_t X_i]+[X_i,[A,X_i]] &=& 0
% [X_i, D_t X_i] &=& jdlkjlsadf \mbox{~}
\mbox{[} X_i , D_t X_i \mbox{]} &=& 0 \ .
\label{eomA}
\eeqa
%% \beqa
%% %\label{eomX} 
%% 0&=& -(D_t)^2 X_i + 
%% [X_j,[X_j,X_i]]
%% + i \alpha \epsilon_{ijk}
%% [X_j,X_k]  \ , \\
%% %i[X_i,\partial_t X_i]+[X_i,[A,X_i]] &=& 0
%% 0 &=& [X_i, D_t X_i]  \ . 
%% \label{eomA}
%% \eeqa
%
There are two types of static solutions.
The first type is given by configurations with
$X_i(t)$ and $A(t)$ being static and diagonal.
The action vanishes identically for such configurations,
and therefore all the diagonal elements are moduli parameters.
The second type of solutions can be represented as
\beq
X_i(t)  = \bigoplus_{I=1}^{s} 
\Bigl( \alpha  \, L_i^{(n_I)} \otimes {\bf 1}_{k_I} \Bigr) \ ,
\quad
A(t) = 
\bigoplus_{I=1}^{s} 
\Bigl( {\bf 1}_{n_I} \otimes \bar{A}^{(I)} \Bigr)
\label{FSA} \ ,
\eeq
where 
%the $n\times n$ Hermitian matrices
$L_i^{(n)}$ represents
%are 
the $n$-dimensional irreducible
representation of the ${\rm SU}(2)$ algebra
$
%\beq
[L_i^{(n)},L_j^{(n)}] = i \, \epsilon_{ijk} \, L_k^{(n)} \ , 
%\eeq
$
%$[L_i^{(n)},L_j^{(n)}] = i\epsilon_{ijk} L_k^{(n)}$,
and the parameters $k_I$ and $n_I$ satisfy
$\sum_{I=1}^{s}n_I \cdot k_I = N \ .$
The $k_I \times k_I$ Hermitian matrices $\bar{A}^{(I)}$ are
arbitrary, and they represent the moduli parameters. 
For this type of classical solutions, 
the action is evaluated as
\beq
S = - \frac{1}{24}\, N \alpha^4 \beta \sum_{I = 1}^s 
({n_I}^3 - n_I ) \, k_I  \ ,
\label{action-FStypeSol}
\eeq
which becomes minimum for $s=1$, $k_1 = 1$, $n_1 = N$.
In this case the solution simply becomes
%In that case the solution simply becomes
\beq
X_i(t)  = \alpha \, L_i^{(N)} \ , \quad
A(t)  = 0 \ ,
\label{sFSX}
\eeq
which represents a single fuzzy sphere 
with the radius $\rho = \frac{1}{2} \, \alpha \sqrt{N^2 -1} $,
since it satisfies
$
%\beq
\sum_{i=1}^3 (X_i)^2 = \rho ^2 \, {\bf 1}_{N} \ .
%\eeq
$
(``Fuzzy'' because of the non-trivial commutation relation
among $X_i$.)

Since the action evaluated for the fuzzy sphere type
solutions (\ref{action-FStypeSol}) is proportional to
$\alpha^4$, it is expected that
the single fuzzy sphere (\ref{sFSX}), which gives the minimum
action among those solutions, dominates the path integral
at sufficiently large $\alpha$.

\section{Perturbative calculation around the fuzzy sphere}
\label{section:pert-calc}

\subsection{Exact effective action and the critical point}
\label{App:one-loop-eff-action}

%%%%%%%%%%%%%%%%%%% gauge fixing %%%%%%%%%%%%%%%%%%%%%%%%%%%%%%% 

In this subsection we calculate the one-loop effective action around 
a configuration $B_i=\kappa L_i^{(N)}$, which
reduces to the single fuzzy sphere solution for $\kappa = \alpha$.
It is known that the effective action 
around a fuzzy sphere configuration
is ``one-loop exact'' in the sense that higher order corrections
vanish in the large-$N$ limit \cite{ANN,0405277,0506205}.
%Therefore, we can obtain the {\em exact} effective action by one-loop
%calculation.
%As a function of $\kappa$, 
%
%The effective action has a local minimum
%if $\alpha$ is larger than a critical value $\alpha_{\rm c}$.
{}From the effective action, we can obtain the critical
coupling $\alpha_{\rm c}$, below which the fuzzy sphere becomes
unstable due to both quantum and thermal fluctuations.
%The obtained critical point $\alpha_{\rm c}$ is also exact 
%to all orders in perturbation theory.

Let us first expand $X_i(t)$ and $A(t)$
around the rescaled single fuzzy sphere $B_i$ as
\be
X_i(t)=B_i+\tilde{X}_i(t) \ , 
\qquad A(t)=0+\tilde{A}(t) \ ,
\label{expandXA}
\ee
where the fields $\tilde{X}_i(t)$ and $\tilde{A}(t)$ represent
the fluctuation.
Since the original action (\ref{action}) 
has a gauge symmetry (\ref{gauge-sym}),
we fix the gauge
by adding the gauge-fixing term and the ghost term as
%to the action $S$ as
\beqa
\label{S_total}
S_{\rm total} &=& S+ S_{\rm g.f.}+S_{\rm gh} \ , \\
S_{\rm g.f.} &=& \frac{1}{2}N  \int\!\! dt\,  
 {\rm tr} \Bigl( \dt A - i[B_i,\tilde{X}_i] \Bigr)^2 \ , \\
S_{\rm gh} &=& N \int\!\! dt\, {\rm tr} 
\Big( \dt \bar{c} \cdot D_t c 
- [B_i,\overline{c}][X_i,c] \Big) \ .
\label{gft}
\end{eqnarray}
Plugging (\ref{expandXA}) into eq.\ (\ref{S_total}),
we obtain $S_{\rm total} = S_0+S_1+S_2+S_3+S_4$, 
where\footnote{We
have omitted a term
$-N \int dt {\rm tr}\left( [B_i,B_j]- 
i\alpha \ijk B_k \right)[\tilde{X}_i,\tilde{X}_j]$
in eq.\ (\ref{Skin}),
which does not contribute to the effective action 
at one loop.
}
\beqa
S_0
&=& \frac{1}{4} \beta N^2(N^2-1)
\left(\frac{1}{2}\kappa^4
-\frac{2}{3}\alpha\kappa^3 \right) \ ,
\label{Scl} \\
S_2
&=& N\int\!\!dt \,{\rm tr} 
\Big\{ \frac{1}{2}\tilde{X}_i
(-\partial^2_t+\kappa^2\mathcal{L}_i^2)\tilde{X}_i
     +\frac{1}{2}\tilde{A}(-\partial^2_t+\kappa^2\mathcal{L}_i^2) 
\tilde{A}+\bar{c}(-\partial^2_t+\kappa^2\mathcal{L}_i^2)c \Big\} \ ,
\label{Skin} \\
S_3 &=& N \int \!\! dt \,{\rm tr}
\Big( -[\tilde{X}_i,\tilde{X}_j][B_i,\tilde{X}_j]
+\frac{2}{3} i\alpha \epsilon_{ijk}
\tilde{X}_i\tilde{X}_j\tilde{X}_k+\bar{c}[B_i,[\tilde{X}_i,c]] \nn \\
&& \hspace{2cm} 
-([\tilde{A},B_i]+i\partial_t\, \tilde{X}_i)[\tilde{A},\tilde{X}_i]
-i\partial_t\,\bar{c}[\tilde{A},c]  \Big) \ ,
\label{Sint}
\eeqa
while the linear term $S_1$ and the quartic term $S_4$ will not be
needed in the following calculation.
In eq.\ (\ref{Skin}), we have introduced the adjoint operation 
$\mathcal{L}_i M \equiv [L_i^{(N)},M]$ 
on an $N\times N$ matrix $M$.
Following the usual procedure, the effective action 
can be calculated as 
$\Gamma(\kappa)
= \Gamma^{(0)}(\kappa) + \Gamma^{(1)}(\kappa)$,
where the classical term is nothing but
$ \Gamma^{(0)}(\kappa) = S_0 $,
and the one-loop term is given as
\be
\Gamma^{(1)}(\kappa) 
= \ln {\rm det}(-\dt^2+\kappa^2 \mathcal{L}_i^2)
\label{1det}
\ee
by performing the Gaussian integration
over the fluctuation fields 
with the quadratic terms (\ref{Skin}).
When taking the determinant in eq.\ (\ref{1det}), we omit
the zero mode corresponding to the constant
mode proportional to the unit matrix.
In order to diagonalize the operator 
$(-\partial^2_t+\kappa^2 \mathcal{L}^2_i)$\,,
we introduce the matrix analog of the spherical harmonics 
$Y_{lm}$ ($0 \le l\le N-1$, $-l \le m \le l$),
which obeys the orthonormal relations
\be
\frac{1}{N} {\rm tr}( Y^\dagger_{lm}Y_{l'm'})
=\delta_{ll'}\delta_{mm'}\,,\quad Y^\dagger_{lm}=(-1)^m Y_{l,-m} \ ,
\label{lower components of Y}
\ee
and has the following properties 
as a representation of the SU(2) algebra
\begin{eqnarray}
\mathcal{L}_3 Y_{lm} &=& m Y_{lm} \ , \nn \\
\mathcal{L}^2_iY_{lm} &=& l(l+1) Y_{lm}\ ,  \\
\mathcal{L}_{\pm} Y_{lm} &=& \sqrt{(l \mp m)(l \pm m+1)}Y_{l,m \pm 1}\ , \nn
\label{eigen equatation for LY}
\end{eqnarray}
where $\mathcal{L}_{\pm}\equiv \mathcal{L}_1 \pm i\mathcal{L}_2$. 
Using 
%these properties and 
the 
%infinite product 
formula
$\prod_{n=1}^{\infty}\left(1+\frac{x^2}{n^2}\right)
=\frac{\sinh \pi x}{\pi x}$, we obtain
\beq
\Gamma^{(1)}(\kappa)
= 2 \sum^{N-1}_{l=1} (2l+1) \ln \left\{ \sinh
\Big(\frac{\beta\kappa}{2}\sqrt{l(l+1)} \Big)\right\}  \ .
\label{lnsinh}
\eeq
Here we have omitted a $\kappa$-independent 
constant,\footnote{This constant
becomes relevant, e.g., when one compares
free energy for different types of vacua \cite{KNY}.}
which is irrelevant for the following analysis.

When we take the large-$N$ limit of the effective
action, we have to scale the parameters $\alpha$, $\beta$
and $\kappa$ in such a way that
the classical term $ \Gamma^{(0)}(\kappa)$
and the one-loop term $\Gamma^{(1)}(\kappa)$
become the same order. This motivates us to
introduce the rescaled parameters
\beq
\tilde{\alpha} \equiv N^{1/3}\alpha \ ,
\quad
\tilde{\beta} \equiv N^{2/3}\beta \ ,
\quad
\tilde{\kappa} \equiv N^{1/3}\kappa \ .
%\quad
%\tilde{T} \equiv \frac{1}{\tilde{\beta} }
%= N^{- \frac{2}{3}} T \ .
\label{rescalded-var}
\eeq
%The effective action divided to $N^2$ becomes order one
%quantity in the large-$N$ limit with fixed $\ta$, $\tk$ and $\tb$.
%Furthermore our model has the property of the one-loop dominance 
%in the large-$N$ limit with fixed $\ta$, $\tk$ and $\tb$ \cite{KTT}. 
%Therefore we can then obtain 
%the exact effective action at large-$N$ within the perturbation theory: 
%\be
%\lim_{N\rightarrow \infty} \Gamma_{\rm all-order}[\tk] 
%= \lim_{N\rightarrow \infty} \Gamma_{\rm 1-loop}[\tk] \,.
%\ee 
The sum over $l$
%mation $\sum_{l=1}^{N-1}$ 
in eq.\ (\ref{lnsinh})
%the one-loop term $\Gamma^{(1)}(\tk)$ 
can be evaluated in the large-$N$ limit with
fixed $\tilde{\beta}$ and $\tilde{\kappa}$.
Thus we obtain the exact effective action as
\beq
\lim_{N\rightarrow \infty} \frac{1}{N^2} \Gamma(\kappa)
= \frac{1}{4}\tb
\Big( \frac{1}{2}\tilde{\kappa}^4
-\frac{2}{3} \tilde{\alpha}\tilde{\kappa}^3 \Big) 
+ \Phi(\tb\tilde{\kappa})  \equiv f(\tk ; \ta , \tb) \ .
\label{Nea}
\eeq
The function $\Phi(x)$ is defined as
\begin{eqnarray}
%\label{one-loopEA0}
\Phi(x) &\equiv& \lim_{N\rightarrow \infty}
\frac{2}{N^2} 
\int_0^N \!\! d\xi \,\, 2\xi \,\,{\rm ln} 
\left\{ {\rm sinh}\left( \frac{x}{2N}\xi \right) \right\}  \nn \\
&=&\frac{1}{3}x -2\ln (1-{\rm e}^{x})+2{\rm ln}
\left({\rm sinh}\,\frac{x}{2}\right)
   -\frac{4}{x}{\rm Li}_2({\rm e}^{x})
+\frac{4}{x^2}{\rm Li}_3({\rm e}^{x})- \frac{4}{x^2} \zeta(3)\ , 
\label{one-loopEA}
\end{eqnarray}
where the polylogarithm function
${\rm Li}_n(z)$ and the Riemann zeta function $\zeta(n)$
are defined, respectively, as
${\rm Li}_n(z)=\sum_{k=1}^{\infty}\frac{z^k}{k^n}$ and 
$\zeta(n)=\sum_{k=1}^{\infty}\frac{1}{k^n}$.

%%%%%% critical point %%%%%%%%%%%%

%As a function of $\kappa$, 
%the effective action has a local minimum
%if $\alpha$ is larger than a critical value $\alpha_c$.

The local minimum of the effective action,
which corresponds to the quantum fuzzy sphere,
can be obtained by solving  
%This moment is the moment that 
%the equation for the derivatived effective action 
%(\ref{Nea}) with $\tilde{\kappa}$ as
\beq
\frac{\del}{\del\tilde{\kappa}}
%\Gamma_{\rm 1-loop}
f(\tk ; \ta , \tb ) = 0 
\label{1ea}
\eeq
with respect to $\tk$ in the region $\tk \sim \ta$.
%, which corresponds to the classical fuzzy sphere.
As we decrease $\ta$, we find that the local minimum
disappears at some critical point $\ta_{\rm c}$, which depends on 
$\tb$.
The critical point $\ta_{\rm c}$ obtained in this way
is plotted against $\tilde{T} \equiv 1/\tb$
in figure \ref{fig:phase-alpha-T}.
%begins to have the real solution for some $\tilde{\alpha}$\,.
%
%In the high temperature limit $\tb \rightarrow 0$,
%the ``sinh'' in eq.\ (\ref{one-loopEA0}) can be replaced by
%its argument, and we obtain the asymptotic behavior
%(\ref{ta_highT}).
%
In particular, the asymptotic behaviors of the critical point 
at the low $\tilde{T}$ and high $\tilde{T}$ limits
are given by
\beq
\ta_{\rm c}  = 
 \left\{
\begin{array}{ll}
9^{1/3}\simeq 2.08 & \mbox{~at~$\tT = 0$}\ , \\
\left(\frac{1024 }{27}\tT \right)^{1/4} 
\simeq 2.48 \tT ^{1/4}  & \mbox{~at~$\tT \gg 1$} \ .
\end{array}
\right.
\label{ta_highT}
\eeq

\subsection{One-loop calculation of observables}
\label{appendix:one-loop}

%eorder to investigate the phase diagram,
In this subsection we
calculate the expectation values of the operators
\begin{eqnarray}
R^2 &\equiv& \frac{1}{N\beta} 
\int_0^{\beta} \!\!dt \, {\rm tr} (X_i)^2 \ , \nn \\
M &\equiv& \frac{2 \, i}{3 \, N\beta} \int_0^{\beta} \!\!dt \, 
\ijk \, {\rm tr}  (X_i X_j X_k) \ ,  \nn \\
F^2  &\equiv& - \frac{1}{N\beta} 
\int_0^{\beta} \!\!dt \, {\rm tr} \Bigl([X_i,X_j]\Bigr)^2 
 \label{DefFMR}
\end{eqnarray}
around the single fuzzy sphere (\ref{sFSX}) at one loop.
Unlike the effective action, the expectation values do have
higher-loop corrections, which shall be obtained in a resummed form
in the next subsection.

Let us decompose the fields into the background and fluctuations as 
in eq.\ (\ref{expandXA}), 
where we set $\kappa = \alpha$ in this subsection.
The expectation value
% of $R^2$ 
$\langle R^2 \rangle$ can be represented as
\beq
  \langle R^2 \rangle
= \frac{1}{N\beta} \int \!\! dt \, {\rm tr}B^2_i 
  +\frac{2}{N\beta} \int \!\! dt \, {\rm tr}B_i\langle \tilde{X}_i(t) \rangle
  +\frac{1}{N\beta} \int \!\! dt \langle  {\rm tr}\tilde{X}^2_i(t) \rangle\ .
\label{Rev0}
\eeq
The first term can be easily evaluated as
\be
\frac{1}{N\beta}\int \!\! dt \, {\rm tr}B^2_i
=\frac{1}{4}\alpha^2(N^2-1) \ . 
\label{1-st term of trx2}
\ee
The second term can be evaluated at one loop
using the cubic terms (\ref{Sint}) as
\beqa
\frac{2}{N\beta} \int \!\! dt \, {\rm tr} B_i\langle \tilde{X}_i(t) \rangle
=& \frac{2}{\beta N} \Big\langle \int \!\! dt \, {\rm tr} 
(B_i \tilde{X}_i(t)) \int \!\! dt' {\rm tr}([\tilde{X}_j(t'),\tilde{X}_k(t')][B_j,\tilde{X}_k(t')]) 
\Big\rangle_0  \nonumber\\
& - \frac{2}{N\beta} \Big\langle \int \!\! dt \, {\rm tr} 
(B_i \tilde{X}_i(t)) \int \!\! dt' {\rm tr}(\bar{c}(t')
[B_j,[\tilde{X}_j(t'),c(t')]]) \Big\rangle_0  \nonumber\\
& + \frac{2}{N\beta} \Big\langle
\int \!\! dt \, {\rm tr} (B_i \tilde{X}_i(t))
\int \!\! dt' {\rm tr}([\tilde{A}(t'),B_j][\tilde{A}(t'),\tilde{X}_j(t')])
\Big\rangle_0 \ ,   
\label{ST of two point function perturbation}
\eeqa
where the symbol
$\langle \hspace{0.2cm}\cdot\hspace{0.2cm} \rangle_0$ 
represents the expectation value using the quadratic terms
(\ref{Skin}) only.
Eq.\ (\ref{ST of two point function perturbation}) 
can be evaluated by using the Wick theorem.
The propagators can be derived from the quadratic terms (\ref{Skin}) as
\begin{eqnarray}
\langle (\tilde{X}_i(t))_{pq}(\tilde{X}_j(t'))_{rs}\rangle_0             
&=& \delta_{ij}\Delta_{pqrs}(t-t') \ ,\\
\langle (\tilde{A}(t))_{pq}(\tilde{A}(t'))_{rs}\rangle_0 
&=& \Delta_{pqrs}(t-t') \ ,\\
\langle (c(t))_{pq}(\bar{c}(t'))_{rs}\rangle_0
&=& \Delta_{pqrs}(t-t') \ ,
\label{propagator}
\end{eqnarray}
where the indices $p,q,r,s$ run over $1,\cdots,N$
and $\Delta_{pqrs}(t-t')$ is defined as
\begin{eqnarray}
\Delta_{pqrs}(t-t')
= \frac{1}{N^2}\sum_{n=-\infty}^{\infty}
\sum_{l=0}^{N-1} {}^{'} \sum_{m=-l}^{l}
\frac{(-1)^m e^{2 \pi in(t-t')/\beta}}
{(2 \pi n/\beta)^2+\alpha^2l (l+1)} 
(Y_{l,-m})_{pq}(Y_{lm})_{r s} \ .
\end{eqnarray} 
The symbol $\sum'$ implies that the zero mode
is omitted by excluding $l=0$ for $n=0$.
Using the formula $\sum_{n=1}^{\infty}\frac{1}{x^2+n^2}
=-\frac{1}{2x}+\frac{\pi}{2x}\coth (x\pi)$,
eq.\ (\ref{ST of two point function perturbation}) 
can be evaluated as
\beq
  \frac{2}{N \beta}\int\!\!dt\,{\rm tr}B_i 
\langle \tilde{X}_i(t) \rangle
%_{\rm 1-loop}
=-\frac{1}{\alpha N^2}\sum_{l=1}^{N-1}(2l+1)\sqrt{l(l+1)}\,
{\rm coth}\Big(\frac{\beta\alpha}{2}\sqrt{l(l+1)}\Big) \ .
\label{2-nd term of trx2}
\eeq
The sum 
%$\sum_{l=1}^{N-1}$ 
over $l$ can be evaluated
at large $N$ as in (\ref{one-loopEA})
for fixed $\ta$ and $\tb$, and it turns out that
(\ref{2-nd term of trx2}) is given by
$-\frac{N}{\alpha} \Phi ' (\tilde{\beta}\tilde{\alpha})$.
%% \bear
%% \Phi(\rho) &\equiv& \frac{1}{N^{3}}\sum_{l=1}^{N-1} (2l+1)\sqrt{l(l+1)} 
%% \coth \left( \frac{\rho}{N}\sqrt{l(l+1)}\right) 
%% \nonumber\\
%% &\simeq& \frac{1}{N^3}\int^N_0 \!\! dx \, 2 x^2 \,
%% \coth\left(\frac{\rho x}{N}\right) 
%% \nonumber\\
%% &=& -\frac{2}{3}+2\ln (1-e^{2\rho})
%% +\frac{2}{\rho}{\rm Li}_2({\rm e}^{2\rho})
%% -\frac{1}{\rho^2}{\rm Li}_3({\rm e}^{2\rho})+\frac{\zeta(3)}{\rho^2} \ ,
%% \label{sum1}
%% \eear
%%where $\rho \equiv \tilde{\alpha}\tilde{\beta}/2$.
%
%and the function $\phi(\rho)$ is given in
%eq.\ (\ref{one-loopEA}).
Since the third term of eq.(\ref{Rev0}) 
is suppressed at large $N$,
%with fixed $\ta$ and $\tb$,
$\langle R^2 \rangle$ is obtained at one loop as
\beq
\lim_{N\rightarrow \infty}\frac{1}{N^{\frac{4}{3}}} 
\langle R^2 \rangle_{\rm 1-loop} 
= \frac{1}{4}\tilde{\alpha}^2
-\frac{1}{\tilde{\alpha}}\Phi'(\tilde{\beta}\tilde{\alpha}) \ .
\label{one-loop result for R2}
\eeq

%%%%%%%%%%%%

The expectation values of $M$ and $F^2$ 
%$\langle M\rangle$ and $\langle F^2\rangle$ 
can be calculated in a similar way, but 
it is much easier to obtain them 
by making use of the fact that
these operators appear in the action (\ref{action}).
The expectation values can therefore be rewritten 
as\footnote{Eq.\ (\ref{Exp F2}) can be derived
by introducing a source term in the action,
and by absorbing it by rescaling the variables as
$t \mapsto  \mu^{-1/3}t$, 
$X_i \mapsto  \mu^{-1/6}X_i$, 
$A \mapsto \mu^{1/3}A$ with an appropriate $\mu$.
Since the integration measure and the kinetic term in 
the action are invariant under this transformation,
the free energy for the action with the source term 
can be obtained by simply rescaling $\alpha$ and $\beta$.}
\begin{eqnarray}
\langle M \rangle &=& \frac{1}{N^2\beta}
\frac{\partial}{\partial \alpha}
W(\alpha,\beta) \ ,
\label{Exp CS}   \\
\langle F^2 \rangle &=& \frac{4}{N^2\beta}
\left( 
- \frac{5}{6} \alpha 
\frac{\partial}{\partial \alpha}
W(\alpha,\beta)
+ \frac{1}{3} \beta
\frac{\partial}{\partial \beta}
W(\alpha,\beta) \right) \ .
\label{Exp F2}
\end{eqnarray}
Here the free energy $W(\alpha,\beta)$ is defined by
\beq
W(\alpha,\beta) = 
-\ln \left(\int[dX][dA]
{\rm e}^{-S} \right) \ ,
\label{lambda free energy}
\eeq
%The action $S$ is gauge fixed as in  eq.(\ref{S+S_gf+S_h})\,.
and at one loop it can be obtained from the effective action 
by simply replacing $\kappa$ by $\alpha$.
In the large-$N$ limit with fixed $\tilde{\alpha}$ and 
$\tilde{\beta}$, we get
\beq
\lim_{N\rightarrow \infty}
\frac{1}{N^2} W_{\rm 1-loop}(\alpha,\beta )
= - \frac{1}{24} \tilde{\beta} \tilde{\alpha} ^4
+ \Phi ( \tilde{\beta} \tilde{\alpha}) \ .
\eeq
%where $\rho \equiv \tilde{\alpha}\tilde{\beta}/2$.
Plugging this into (\ref{Exp CS}) and (\ref{Exp F2}),
we obtain
\beqa
\lim_{N\rightarrow \infty}\frac{1}{N} \langle M \rangle_{\rm 1-loop}
&=& -\frac{1}{6}\tilde{\alpha}^3
+ \Phi'(\tilde{\beta}\tilde{\alpha})  \ ,
\label{one-loop result for M}\\
\lim_{N \rightarrow \infty}
\frac{1}{\,N^{\frac{2}{3}}}\langle F^2  \rangle_{\rm 1-loop}
&=& \frac{1}{2}\tilde{\alpha}^4
- 2 \ta \Phi ' (\tilde{\beta}\tilde{\alpha}) \ . 
\label{one-loop result for F2}
\eeqa

\subsection{All order calculation of observables}
\label{appendix:all-order}

In this subsection
we exploit the fact that the effective action
is saturated at one loop in the large-$N$ limit, 
and calculate the expectation values 
of the operators $R^2$, $M$ and $F^2$
to all orders in perturbation theory.
The crucial point here is that the free energy and the effective
action are related to each other by the Legendre transformation.
Therefore, we can obtain the free energy
by evaluating the effective action at its local minimum.
Since the expectation values can be obtained
by differentiating the free energy (for an action
including an additional source term if the operator
does not exist in the original action),
we can obtain the all order results for 
the expectation values in the large-$N$ limit.
%The crucial point is that 
%the free energy and the effective action 
%are related to each other by the Legendre transformation. Namely 
%the free energy can be obtained by evaluating 
%the effective action at its extremum. 
%This extremum point corresponds to a true vacuum 
%%which is 
%including quantum effect. 
%Further, in our study, 
%the effective action enjoys the one-loop dominance in the large-$N$
%limit as well as ref.\cite{ANN}. 
%Therefore, by evaluating the one-loop effective action at its extremum, 
%we can obtain the free energy and hence the observables 
%to all order in the large-$N$ limit.
%
%The resulting prescription to obtain 
%the all-order results
%is to consider only the terms in the one-loop result
This amounts to \cite{ANN}
keeping only the terms in the one-loop result
that come from 1PI diagrams, and 
replacing $\tilde{\alpha}$
by the solution to eq.\ (\ref{1ea}),
which we denote as $\tk_{0}$ in what follows.
Since the one-loop contributions to 
$\langle M \rangle$ and
$\langle R^2 \rangle$ come only from 1PR diagrams,
the corresponding all order results are readily obtained
from the classical results
by replacing $\tilde{\alpha}$ by $\tk_{0}$ as
\bear
\label{all-order result1}
\frac{1}{N^{\frac{4}{3}}}\langle R^2 \rangle_{\rm all-order} 
&=& \frac{1}{4} (\tilde{\kappa}_{0})^2 \ , 
\\%\,,\quad
\frac{1}{N}\langle {M} \rangle_{\rm all-order}
&=& -\frac{1}{6} (\tilde{\kappa}_{0})^3 \ .
\label{all-order result2}
\eear

%As a concrete observable let us consider $M$\,, $R^2$ and $F^2$\,. 
%%What has to be noticed is that, in the representation 
%For $M$ and $R^2$\,, the contributions 
%from one-loop part are totally given by the tadpole diagram. 
%This is a 1PR diagram. Thus it turns out that 
%we should consider here to obtain all-order result 
%one-loop expectation values of $M$ and $R^2$ are only given 
%by classical part, 
%%\begin{align}
%\bear
%\frac{1}{N} \langle M \rangle_{\rm 1-loop}\Big|_{\rm 1PI} 
%&=& -\frac{\tilde{\alpha}^3}{6} 
%\label{1PI1} \\%\,,\quad
%%\frac{1}{N^{\frac{2}{3}}}\langle F^2 \rangle=  \frac{\tilde{\alpha}^4}{8}\,,\quad
%\frac{1}{N^{\frac{4}{3}}} \langle R^2 \rangle_{\rm 1-loop}\Big|_{\rm 1PI} 
%&=& \frac{\tilde{\alpha}^2}{4}\,.
%\label{1PI2}
%\eear
%%\end{align} 
%By plugging the solution $\tilde{\kappa}_{0}$ 
%%as in (\ref{kaapa solution}) 
%into $\tilde{\alpha}$ of (\ref{1PI1}) and (\ref{1PI2}), 
%we can obtain the all-order results as follows: 
%%\begin{align}
%\bear
%\frac{1}{N}\langle {M} \rangle_{\rm all-order}
%&=& -\frac{\tilde{\kappa}_{0}^3}{6} \\%\,,\quad
%\frac{1}{N^{\frac{4}{3}}}\langle R^2 \rangle_{\rm all-order} 
%&=& \frac{\tilde{\kappa}_{0}^2}{4}\,.
%\label{all-order result}
%\eear
%%\end{align}

Let us next consider $\langle F^2 \rangle $.
Since the one-loop contribution to $\langle F^2 \rangle $
includes both 1PI diagrams and 1PR diagrams,
it is easier to obtain the all order result by
using the relation (\ref{Exp F2}).
As explained above, 
the free energy is given
to all orders in perturbation theory as
%we obtain
\beq
\lim_{N\rightarrow \infty}\frac{1}{N^2}
W_{\rm all-order}(\alpha , \beta)
=  f(\tk_{0} ; \ta , \tb ) \ . 
\eeq
When we differentiate $W_{\rm all-order}(\alpha , \beta)$
with respect to $\alpha$ and $\beta$, 
%in eq.\ (\ref{Exp F2}), 
we have to take into account that $\tk_{0}$ depends 
on $\ta$ and $\tb$. Thus we obtain
\bear
& & \lim _{N \rightarrow \infty}
\frac{1}{N^{\frac{2}{3}}} \langle F^2 \rangle_{\rm all-order} \nn \\
&=& \frac{4}{\tb}\left(-\frac{5}{6} \ta \frac{d}{d\ta}
+\frac{1}{3}\tb \frac{d}{d\tb} \right) 
f (\tk_{0} ; \ta,\tb)
%\Gamma_{\rm 1-loop}[\ta,\tb,\tk_{0}(\ta,\tb)]
%\Big|_{\tk=\tk_{0}} 
\nn \\
&=& \left. \frac{4}{\tb} 
\left\{ -\frac{5}{6} \ta \left(\frac{\del}{\del\ta}
+\mathcal{A}\frac{\del}{\del\tk}\right)
+\frac{1}{3} \tb \left(\frac{\del}{\del\tb}+\mathcal{B}
\frac{\del}{\del\tk}\right) \right\} 
f (\tk ; \ta,\tb) \right|_{\tk=\tk_{0}}  \ ,
%\Gamma_{\rm 1-loop}[\ta,\tb,\tk]\Big|_{\tk=\tk_{0}} 
\label{all-F2}
\eear
where the coefficients $\mathcal{A}$ and $\mathcal{B}$
are given as
\beqa
\mathcal{A} &\equiv& \frac{\del\tk_{0}}{\del\ta}
= 
- \left. \frac{\del^2 f(\tk ; \ta,\tb)}{\del\tk\del\ta}
\right|_{\tk=\tk_{0}} 
 \left( \left. \frac{\del^2 f(\tk ; \ta,\tb) }{\del^2\tk} 
\right|_{\tk=\tk_{0} }\right)^{-1}  \ , \\
\mathcal{B} &\equiv& \frac{\del\tk_{0}}{\del\tb}  
= 
- \left. \frac{\del^2 f(\tk ; \ta,\tb)}{\del\tk\del\tb}
\right|_{\tk=\tk_{0}} 
 \left( \left. \frac{\del^2 f(\tk ;\ta,\tb) }{\del^2\tk} 
\right|_{\tk=\tk_{0} }\right)^{-1}  \ . 
\label{defXY}
\eeqa 
%Here, we rescale in a similar way to eq.(\ref{rescale alpha and beta})\,.
%In doing so, we drop the infinite contribution by hand.
%As for $X$ and $Y$\, 

\section{Comparison with Monte Carlo results}
\label{sec:boundary}

In this section we compare the all order results obtained
in the previous section with the results of Monte 
Carlo simulation taking the single fuzzy sphere (\ref{sFSX}) 
as the initial configuration.
The lattice formulation and the algorithm 
used for simulating the model (\ref{action})
is the same as
in ref.\ \cite{bbfss}.
The lattice spacing $a$ and the number of sites $N_{\rm t}$
in the Euclidean time direction obey the relation
$N_{\rm t} a = \beta$.
We have chosen these lattice parameters so that our results 
represent the continuum limit 
with sufficiently good accuracy.\footnote{More precisely, 
the lattice parameters are chosen to satisfy
both $a \le \epsilon$ and $N_{\rm t}\ge 10$ at any temperature,
where $\epsilon=0.02$ is used for figures \ref{fig:Phase_Trans_lower}
and \ref{Fig_N16_alp_3o0} (except for the right bottom panel),
% \ref{Fig_N16s_alp_0},
% \ref{Fig_N16s_alp_3o8},
and $\epsilon=0.05$ otherwise.
See ref.\ \cite{Kawahara:2005an} for an analysis on finite
lattice spacing effects in a related model.
}
%2\ref{fig:phase-alpha-T},
%4{Fig_N16_alp_0o0_PL}
%5{Fig_N16_alp_0o0}

\subsection{Boundary of the fuzzy sphere phase}

\FIGURE{
\epsfig{file=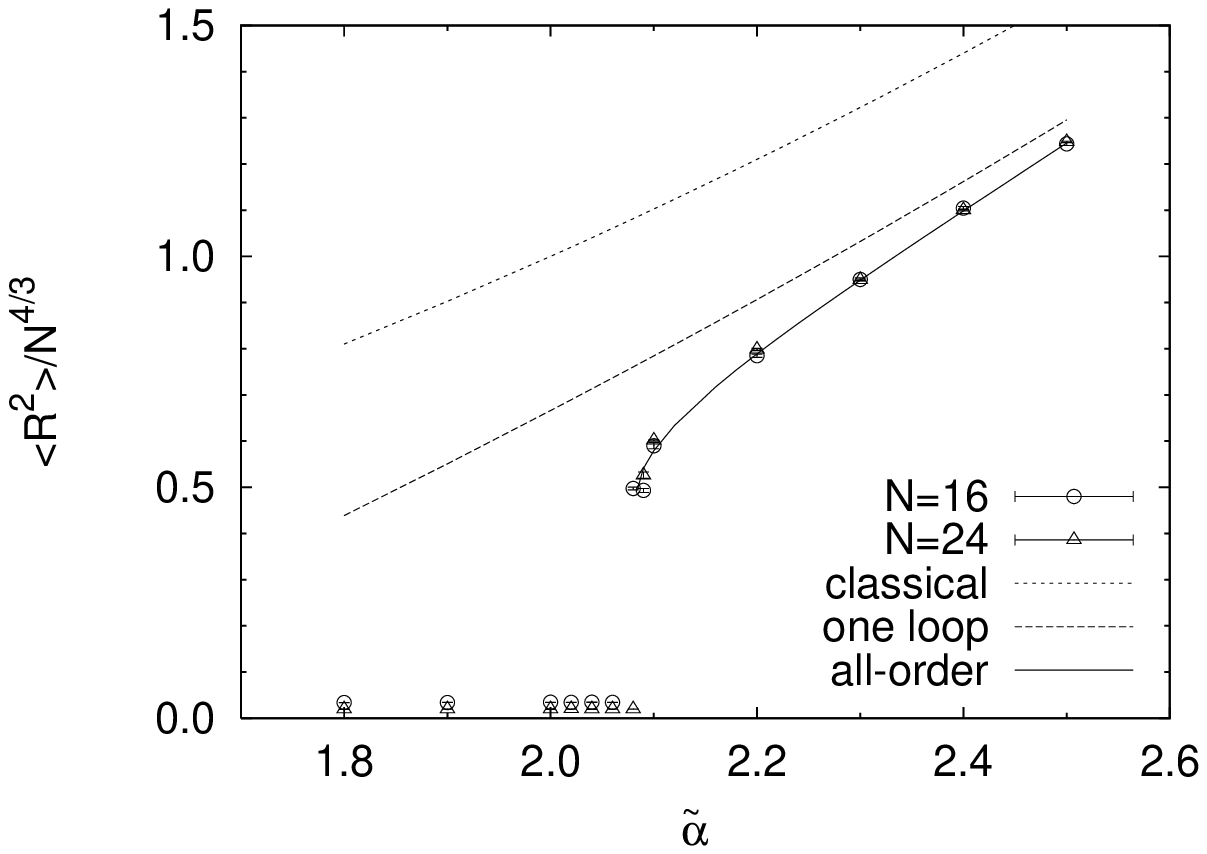,width=7.4cm}
\epsfig{file=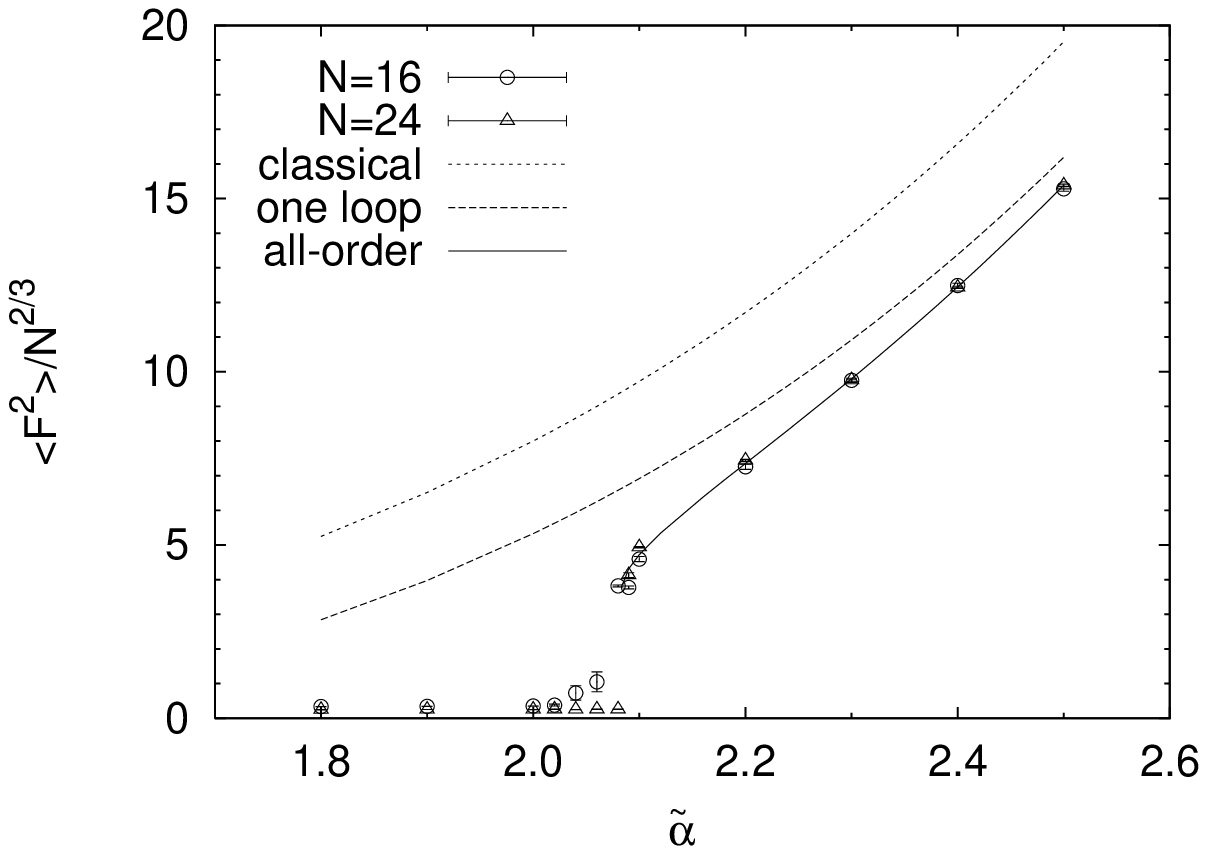,width=7.4cm}
\epsfig{file=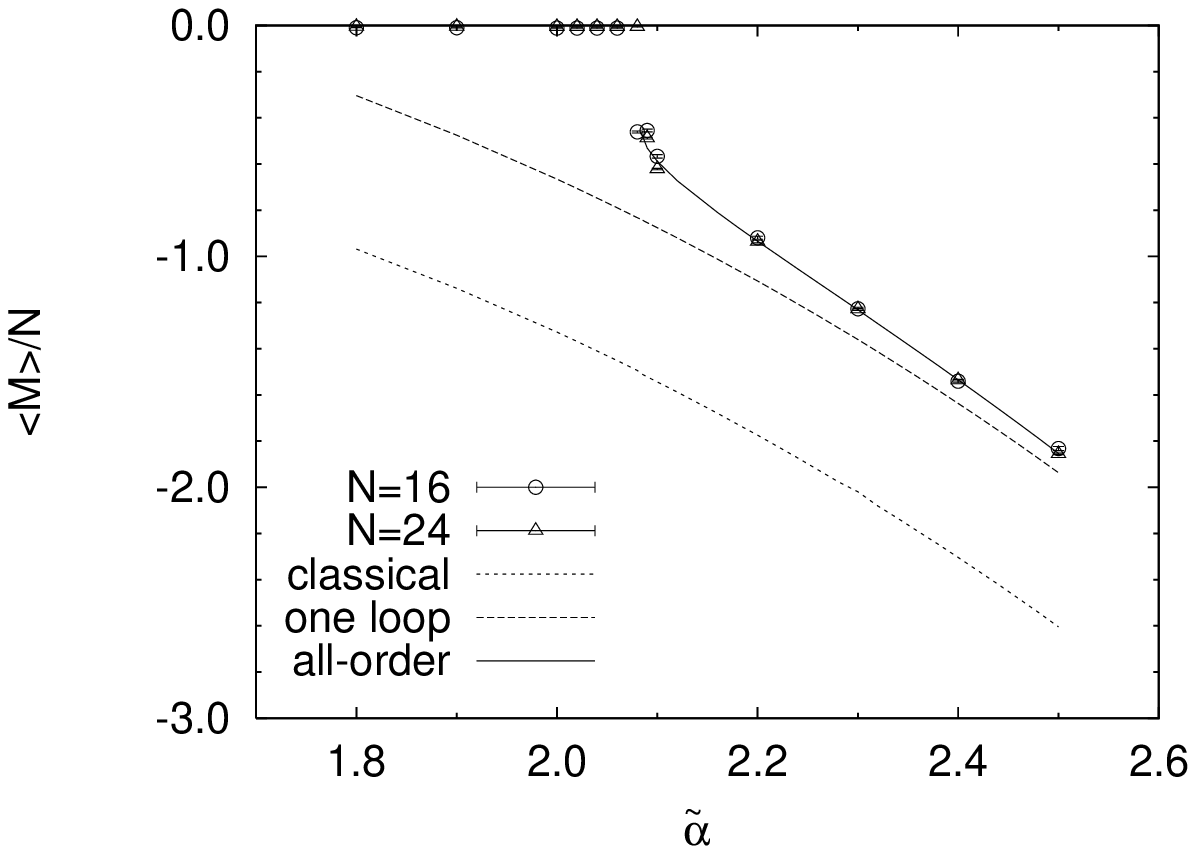,width=7.4cm}
\epsfig{file=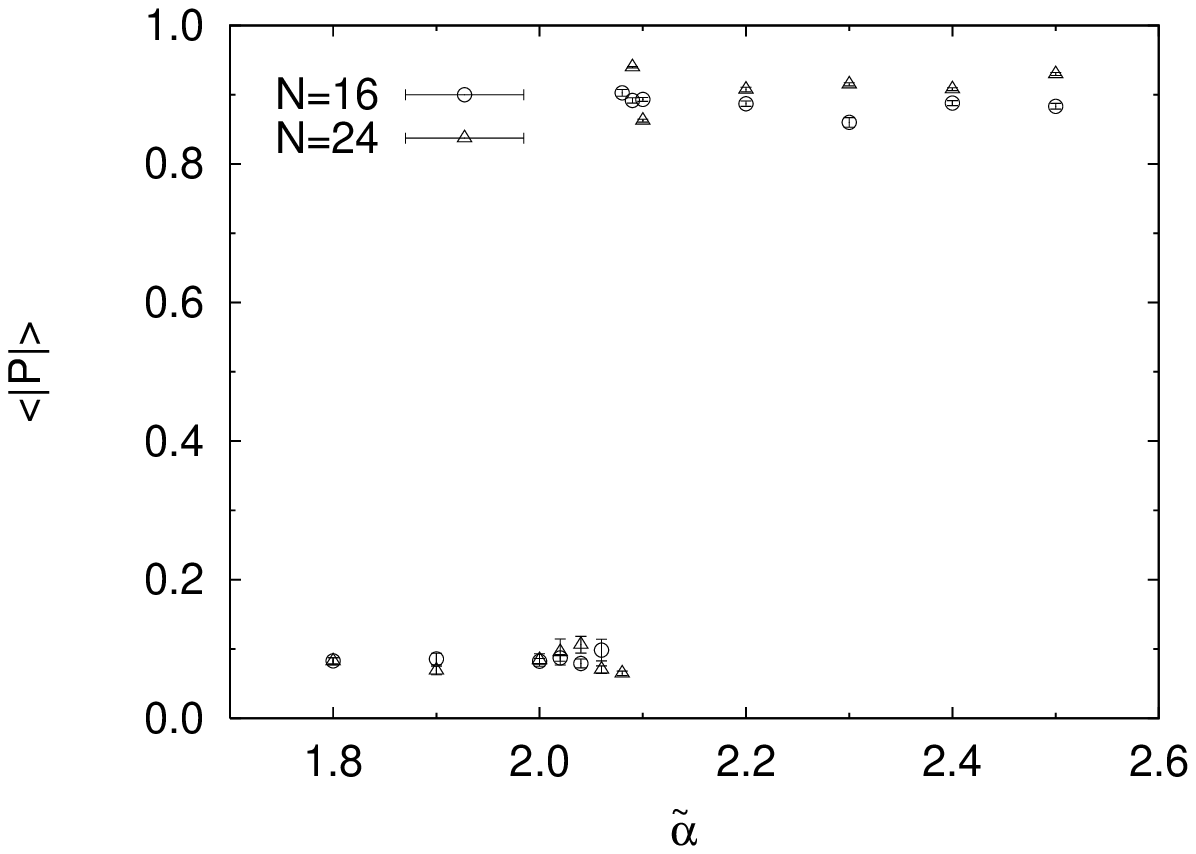,width=7.4cm}%
\caption{The observables
$\langle R^2 \rangle / N^{4/3}$,
$\langle F^2 \rangle / N^{2/3}$,
$\langle M \rangle /N$ and $\langle |P| \rangle$ 
%Various observables 
are plotted against ${\tilde \alpha}$
for $\tilde{T}=0.1$.
% ($T=0.633$) and at $N=16$.
%Monte Carlo simulation has been performed with
%the lattice parameters $(N_t=79, \, a=0.02)$.
The dotted, dashed and solid lines represent the classical,
one-loop and all order results, respectively.
\label{fig:Phase_Trans_lower}
}}

\FIGURE{
\epsfig{file=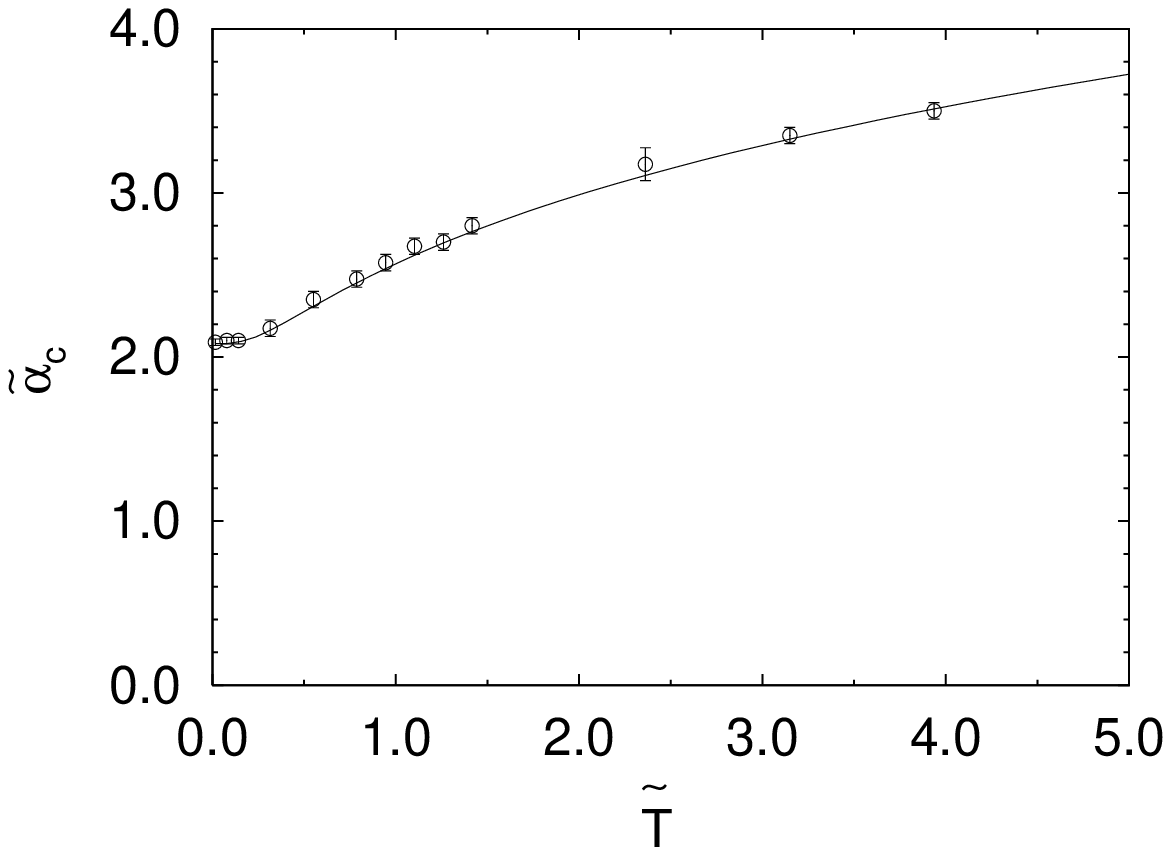, width=8.6cm}
\caption{
The critical $\tilde{\alpha}$, below which the fuzzy sphere
becomes unstable, is plotted against $\tT$.
The open circles represent the results obtained
by Monte Carlo simulation for $N=16$.
%(The lattice spacing is taken to be $a=0.05$ for $T<2.0$,
%and $a=1/(10T)$ for $T \geq 2.0$.)
The solid line represents the result obtained
from the one-loop effective action in the large-$N$ limit.
%The dashed line represents the result (\ref{talpha in HTR})
%obtained from the dimensionally reduced model, which provides
%the high temperature description.
}
\label{fig:phase-alpha-T}}

Let us first investigate how the observables 
(\ref{DefFMR}) behave as a function of $\alpha$.
This, in particular, allows us to 
determine the critical $\alpha$, below which
the single fuzzy sphere becomes unstable.
%The all order results are obtained as
%(\ref{all-order result1}), (\ref{all-order result2}) 
%and (\ref{all-F2}).

In figure \ref{fig:Phase_Trans_lower} we plot the expectation
values 
$\langle R^2 \rangle / N^{4/3}$,
$\langle F^2 \rangle / N^{2/3}$,
$\langle M \rangle /N$
against $\ta$ for fixed $\tilde{T}$ close to $\tilde{T}=0$.
%In particular we have introduced
Monte Carlo results 
show a discontinuity at $\ta \sim 2.1$,
which agrees with the result (\ref{ta_highT}) at $\tT=0$.
Above the critical point,
Monte Carlo results for $N=16,24$ lie on top of each other
as expected from perturbation theory,
and they agree very well with 
the all order results
given by (\ref{all-order result1}), (\ref{all-order result2}) 
and (\ref{all-F2}).
We have also plotted the classical results and the one-loop results
for comparison.
It clearly demonstrate
the existence of higher-loop corrections,
which are included in the all order results.

%each other with this scaling.
%We observe that Monte Carlo data for different $N$ lie on top of
%each other with this scaling.

In Monte Carlo simulation
we also calculate the Polyakov line
\beq
P \equiv \frac{1}{N}{\rm tr} \, 
\mathcal{P} \exp\left(i \int_0^{\beta} \!\!dt A(t) \right)\ ,
\label{pol-def}
\eeq
where the symbol $\mathcal{P}\exp$ represents the path-ordered
exponential.
Results for $\langle |P| \rangle$ 
are shown
in the right bottom panel of figure \ref{fig:Phase_Trans_lower}.
We observe a gap at the same $\ta$ as the other observables.
The properties of the Polyakov line
% in the large $N$ limit
will be discussed later in more detail.
%\footnote{If one increases
%$N$ further with the same $\tT$, the Polyakov line 
%$\langle |P| \rangle$ will be non-zero on both sides of the
%further in subsequent sections.

{}From Monte Carlo simulations at various $\tilde{T}$,
we obtain the critical $\ta$ as a function of $\tilde{T}$,
which is plotted in figure \ref{fig:phase-alpha-T}.
%where we also plot the critical $\ta$ obtained 
%from the perturbation theory 
%in section \ref{App:one-loop-eff-action}.
%As we discuss in Appendix \ref{App:Derivation of the effective action},
We observe perfect agreement
with the results obtained from the one-loop effective action
in the large-$N$ limit.
% in section \ref{App:one-loop-eff-action}.
This confirms that the effective action
is indeed saturated at one loop.
%This property play the crucial role in obtaining
%all-order results for the observables essentially
%from the one-loop calculation. (See Appendix \ref{appendix:all-order}.)

We call the region above the critical line the fuzzy sphere phase,
and the region below the critical line the Yang-Mills phase,
following the terminology used in ref.\ \cite{0401038}.
The phase transition between the fuzzy sphere phase
and the Yang-Mills phase continues to be
of first order at any temperature,
judging from the existence of discontinuity.
%The first order nature of the phase transition will be discussed
%in more detail in the high temperature limit
%in section \ref{sec:dim-red}.
In section \ref{sec:YM-Hage}
we will see that
the Yang-Mills phase is further divided into two phases
by the Hagedorn transition.

\subsection{Temperature dependence of observables}

Next we
investigate the temperature dependence of observables.
In figure \ref{Fig_N16_alp_3o0} we plot
the expectation values
$\langle R^2 \rangle / N^{4/3}$,
$\langle F^2 \rangle / N^{2/3}$,
$\langle M \rangle /N$
against $\tilde{T}$ for $\ta = 3.0$.
There is a gap at $\tT\simeq2.0$,
as expected from figure \ref{fig:phase-alpha-T}.
The all order results reproduce the
$\tT$ dependence of the observables very well 
below the critical $\tT$.
Thermal effects tend to shift
%reduce the radius of the fuzzy sphere.
the observables towards the values above the 
critical temperature.
%% We have also confirmed that other observables 
%% below the critical $\tT$ are nicely reproduced by the
%% all order results and they show nontrivial 
%% $\tT$ dependence towards typical values
%% in the Yang-Mills phase.
%
%, and that the 
%Monte Carlo results below the critical $\tT$ 
%are reproduced nicely by the all order results as well.

%% \FIGURE{
%% %\epsfig{file=N16_alp3o0_PT.eps, width=7.4cm}
%% \epsfig{file=N16_alp3o0_CS.eps, width=7.4cm}
%% \epsfig{file=N16_alp3o0_X2.eps, width=7.4cm}
%% \caption{
%% $\langle M \rangle/N$  and 
%% $\langle R^2 \rangle/N^{\frac{4}{3}}$
%% %$\langle F^2 \rangle/N^{\frac{2}{3}}$ 
%% are plotted against $\tilde{T}\,
%% (\,\textrm{for } T \leq 5.0 \,, a=0.02,\,N_t=1/(0.02T) 
%% \textrm{ and for } T > 5.0 \,, N_t=10,\,a=1/(10T))$   
%% for $\tilde{\alpha}=3.0$ and $N=16$\,. 
%% The dotted, dashed, solid lines represent the classical,
%% one-loop, all-order results, respectively.}
%% \label{Fig_N=16_alp=3o0}}

\FIGURE{
\epsfig{file=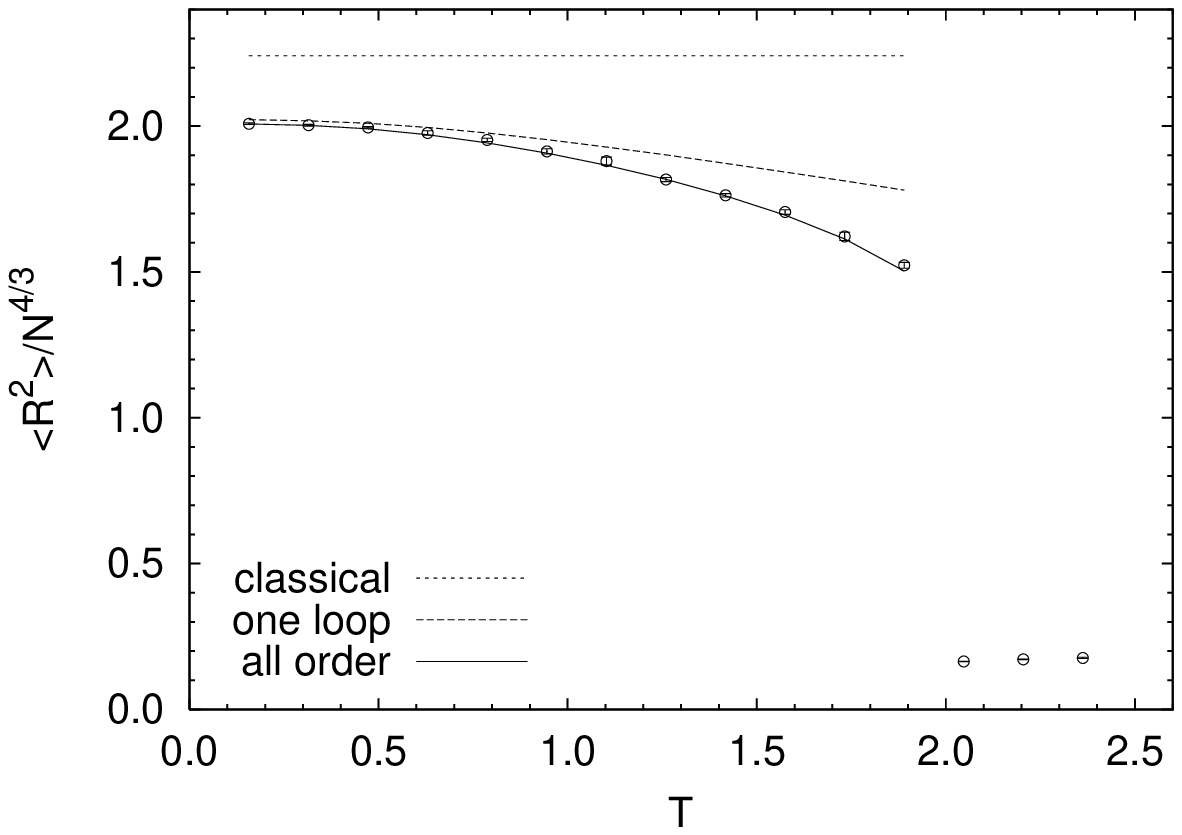, width=7.4cm}
\epsfig{file=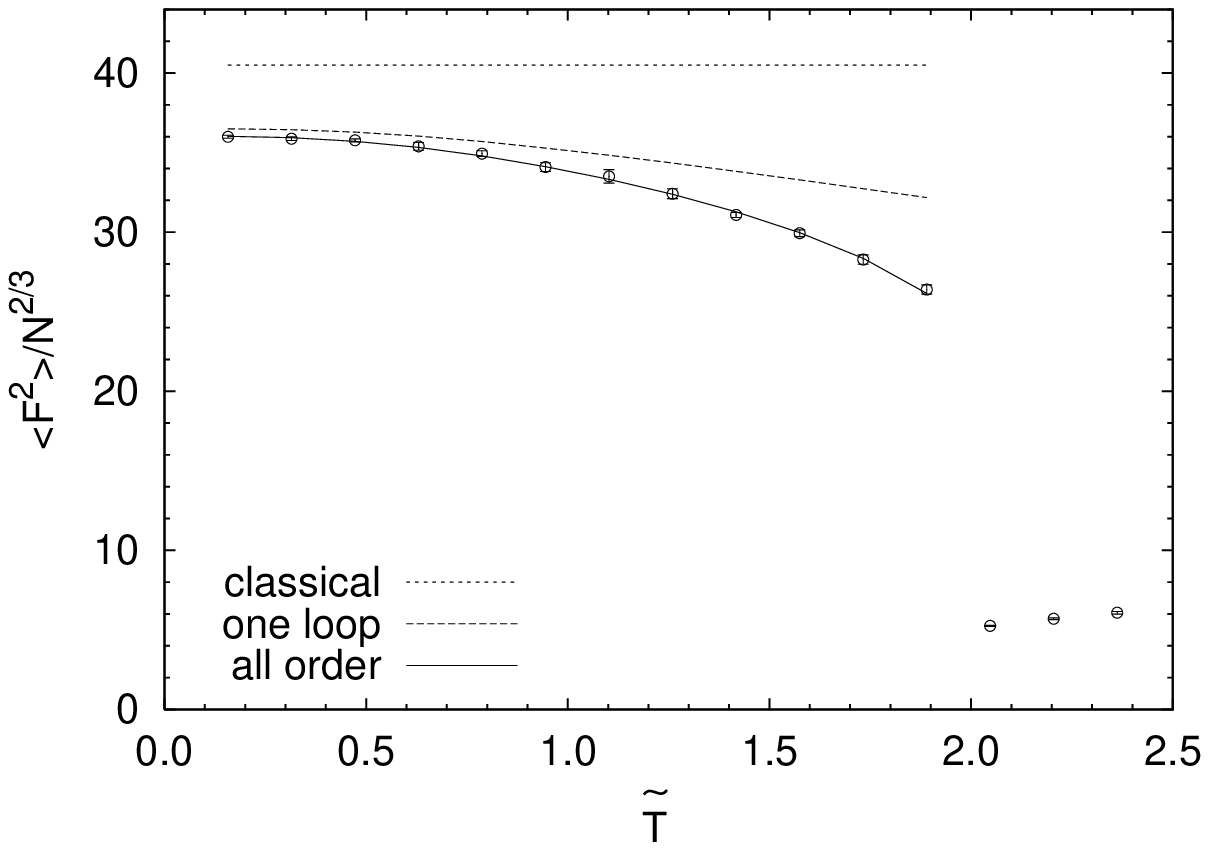, width=7.4cm}
\epsfig{file=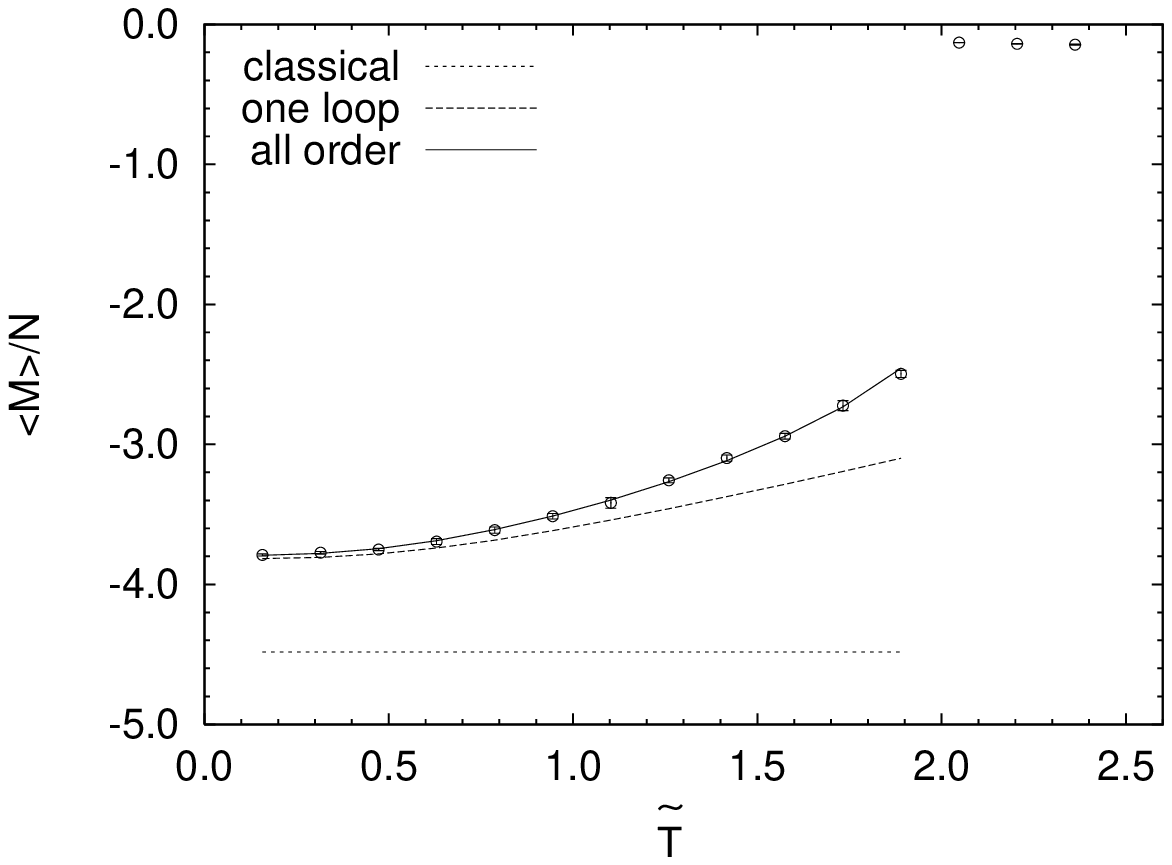, width=7.4cm}
\epsfig{file=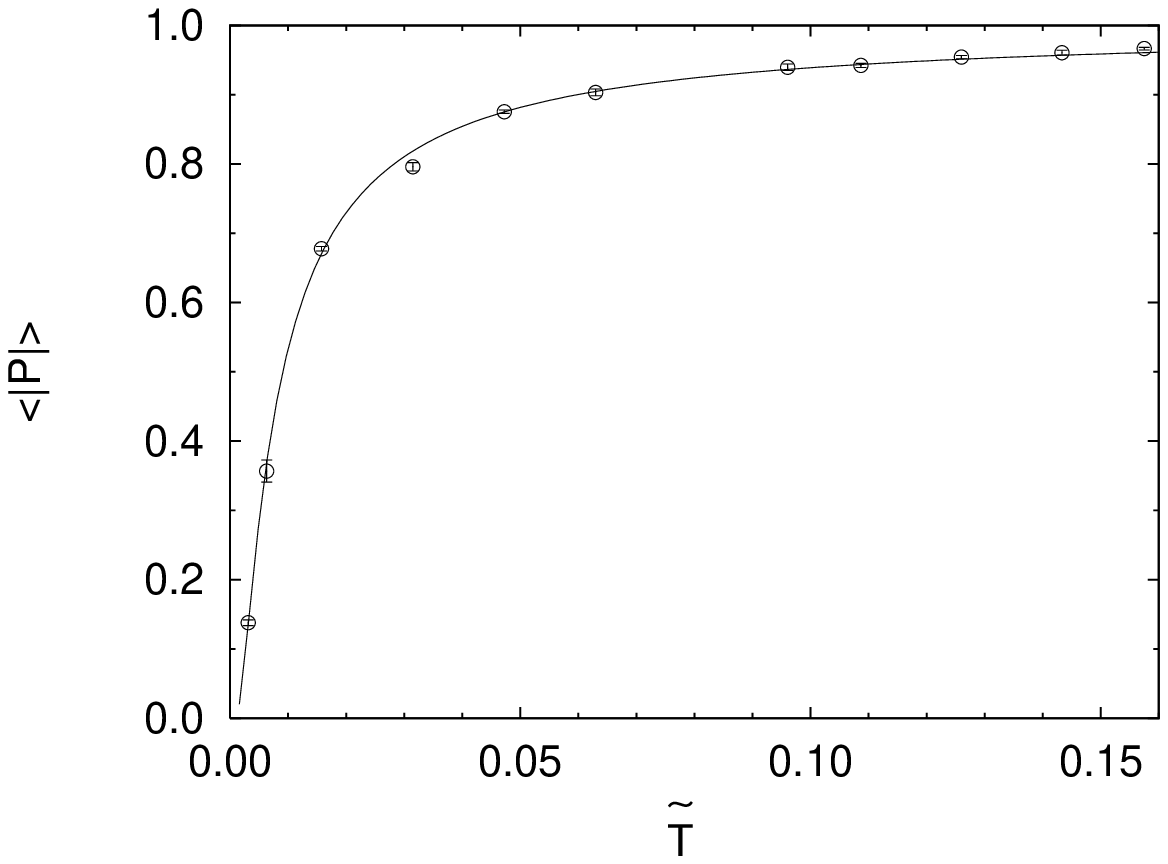, width=7.4cm}
\caption{
The expectation values
$\langle R^2 \rangle / N^{4/3}$,
$\langle F^2 \rangle / N^{2/3}$,
$\langle M \rangle /N$
are plotted against $\tilde{T}$
%(\,\textrm{for } T \leq 5.0 \,, a=0.02,\,N_t=1/(0.02T) 
%\textrm{ and for } T > 5.0 \,, N_t=10,\,a=1/(10T))$   
for $\tilde{\alpha}=3.0$ and $N=16$\,. 
The dotted, dashed, solid lines represent the classical,
one-loop, all order results, respectively.
In the right bottom panel we plot
$\langle |P| \rangle$ 
against $\tilde{T}$, focusing on the small $\tT$ region, 
%($a=0.05$, $N_t=1/(0.05T)$)
for $\tilde{\alpha}=3.0$ and $N=16$\,. 
The solid line represents a fit to eq.\
(\ref{qqbar}).
% with $c =0.0063$\,.
}
\label{Fig_N16_alp_3o0}}

In the right bottom panel of
figure \ref{Fig_N16_alp_3o0},
we plot the Polyakov line $\langle |P| \rangle$ 
as a function of $\tT$. 
We have magnified the small $\tT$ region
in order to see how the Polyakov line 
decreases as $\tT$ approaches 0.
(Note that the scale of $\tT$ in this plot is 
an order of magnitude smaller
than in other plots in figure \ref{Fig_N16_alp_3o0}.)
Our results can be nicely fitted to the behavior 
\be
\langle |P| \rangle=\exp\Big( -\frac{c}{\tilde{T}} \Big) \ ,
\label{qqbar}
\ee
which suggests that the system is in the ``deconfined phase''.
The fitting parameter $c=0.0063$ corresponds 
to the energy increase caused by a single heavy ``quark''.
%increase caused by the existence of a single heavy quark
{}From this figure we 
conclude that the center symmetry is always broken
at $\tilde{T}\neq 0$.
This statement needs some care, however.
See footnote \ref{fuzzy-EK}.

%% Note that, in the fuzzy sphere phase,
%% we have been considering super high temperature
%% $T = {\rm O}(N^{2/3})$ in the large-$N$ limit.
%% If we fix $T$ rather than $\tT$ in the 
%% $N\rightarrow \infty$ limit, we obtain
%% results, which have no dependence on $T$.
%% %The results would be nothing but what we have obtained
%% %at $\tT =0$.
%% %In particular, 
%% %if we took the large-$N$ limit at finite $T$,
%% In particular, the Polyakov line vanishes identically, 
%% meaning that the center symmetry is unbroken for arbitrary $T$.

%% \FIGURE{
%% \epsfig{file=N16_alp3o0_PL.eps, width=7.4cm}
%% \caption{$\langle P \rangle$ is plotted 
%% against $\tilde{T}(\equiv TN^{-2/3})\,
%% (a=0.05\,,N_t=1/(0.05T))$  
%% for $\tilde{\alpha}=3.0$ and $N=16$\,. 
%% The solid line represents fitting line 
%% by taking $\tilde{F}_q =0.04N^{\frac{2}{3}}$\,.}
%% \label{Fig_N=16_alp=3o0-new}}

\section{Hagedorn transition in the Yang-Mills phase}
\label{sec:YM-Hage}

In this section we investigate the properties of 
the Yang-Mills phase.
Perturbation theory is not applicable here, but
Monte Carlo simulation continues to be a reliable
method. 

\FIGURE{
       \epsfig{file=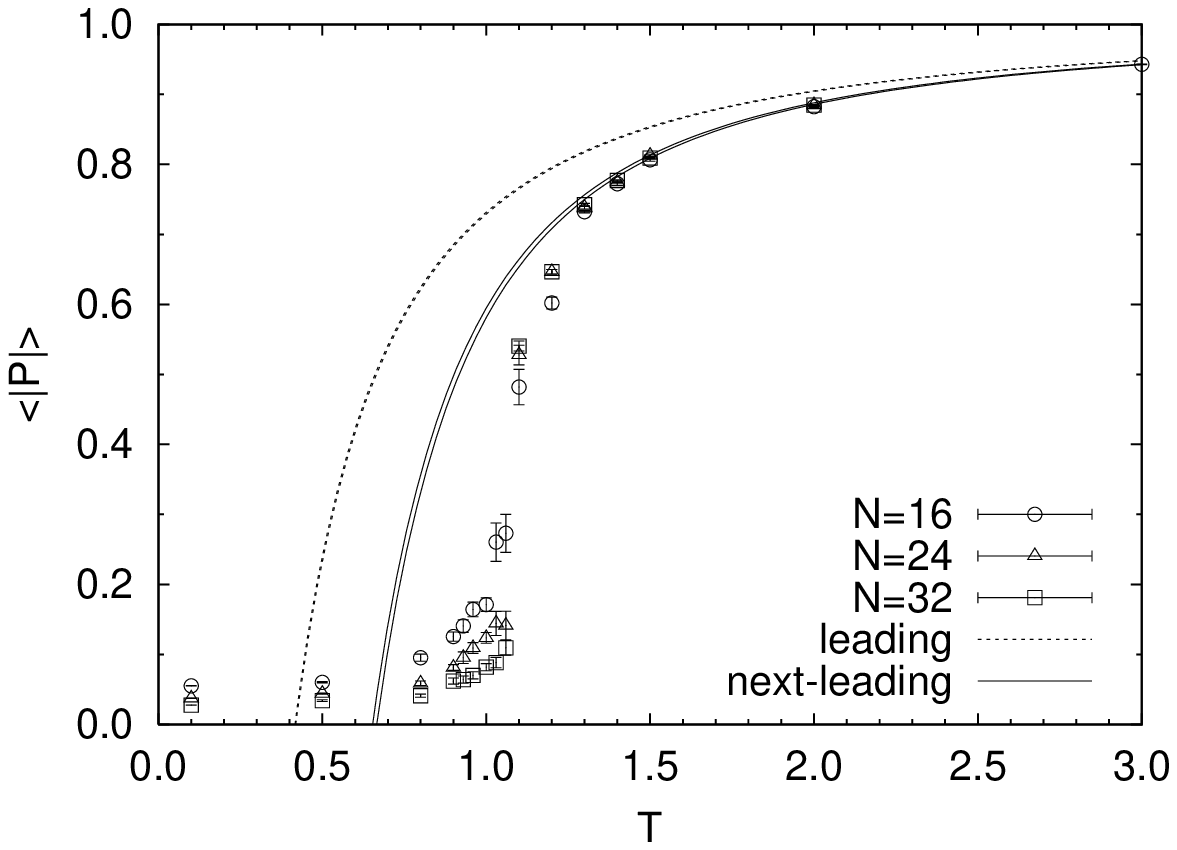, width=7.4cm}
\caption{
The Polyakov line $\langle |P| \rangle$ is plotted against $T$
%$(a=0.05,\,N_t=1/(0.05T))$  
for $\alpha=0.0$ and $N=16,24,32$\,.
The dashed line represents the result obtained
by eq.\ (\ref{polDR}) in the high $T$ limit.
%for the dimensionally reduced model (\ref{DRmodel}). 
The solid lines represent the result
including the next-leading order terms \cite{highT}.
}
\label{Fig_N16_alp_0o0_PL}}  

In figure \ref{Fig_N16_alp_0o0_PL}
the Polyakov line $\langle |P| \rangle$ is
plotted against $T$.
We find that it changes very rapidly
at the temperature $T \sim 1.1$, 
which we denote as $T_{\rm H}$.
Above $T_{\rm H}$,
the data are clearly nonzero, and
they have little dependence on $N$.
Below $T_{\rm H}$, the data are
consistent with
$\langle |P| \rangle$ decreasing as
$1/N$ at large $N$.
Thus our data suggest that the center symmetry 
is spontaneously broken at $T > T_{\rm H}$.
This transition can be interpreted as the Hagedorn 
transition \cite{Aharony,FSS}, and the critical temperature
$T_{\rm H}$ is referred to as the Hagedorn temperature
in what follows.
The value of $T_{\rm H}$ is close to 
the result
$T_{\rm H} \simeq \lambda^{1/3}$
obtained in ref.\ \cite{latticeBFSS}\footnote{Note, however, 
that the lattice model studied in ref.\ \cite{latticeBFSS}
is written in terms of unitary matrices $U_i(t)$ 
instead of Hermitian matrices $X_i(t)$, and 
it agrees with our model only after replacing
$U_i(t)$ by $\exp (ia X_i(t))$ and truncating the action
at the leading order in the lattice spacing $a$.
Let us also note that an analogous model with
9 (instead of 3) Hermitian matrices has been studied by Monte Carlo 
simulation \cite{Aharony:2004ig,bbfss} from different
motivations. In that case the phase transition
occurs at $T \sim 0.9$, which is slightly lower
than the present model.}, where $\lambda$
is the 't Hooft coupling constant, which is set to unity
in our analysis. (See footnote \ref{footnote:tHooft}.)

\FIGURE{
       \epsfig{file=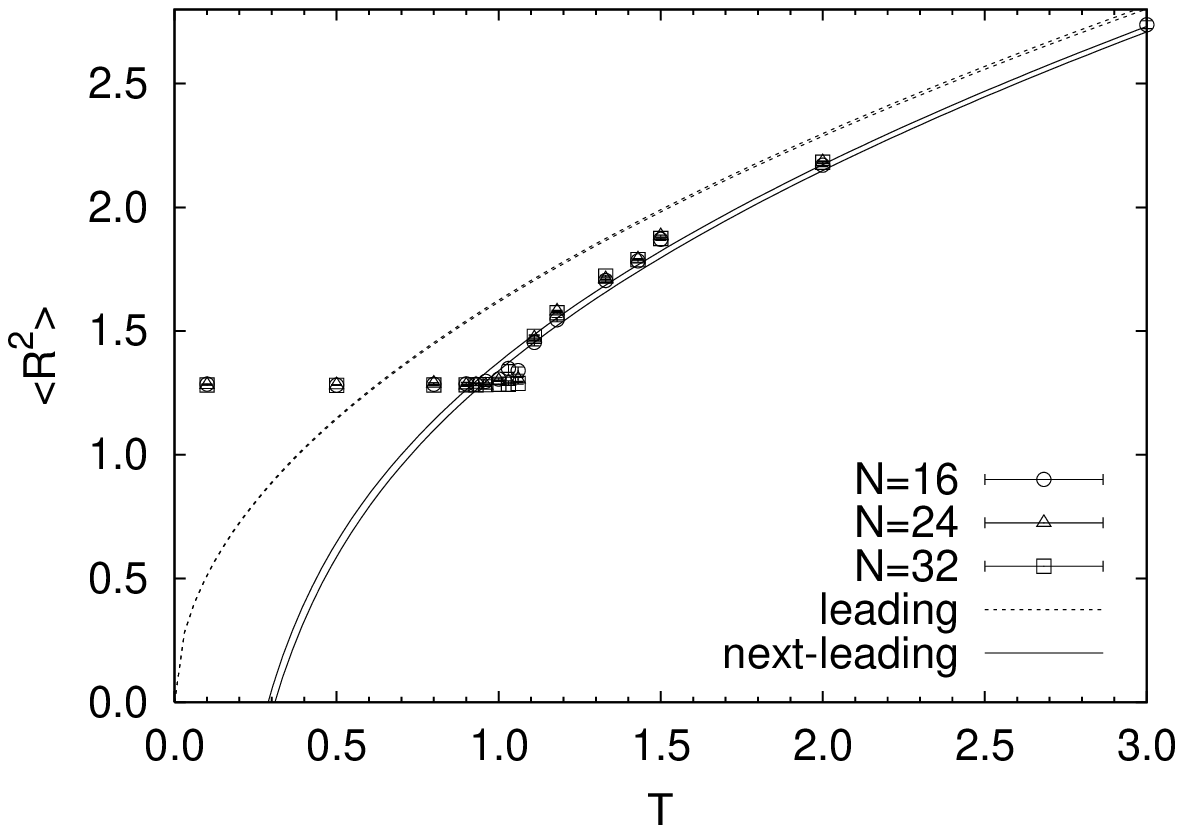, width=7.4cm}
       \epsfig{file=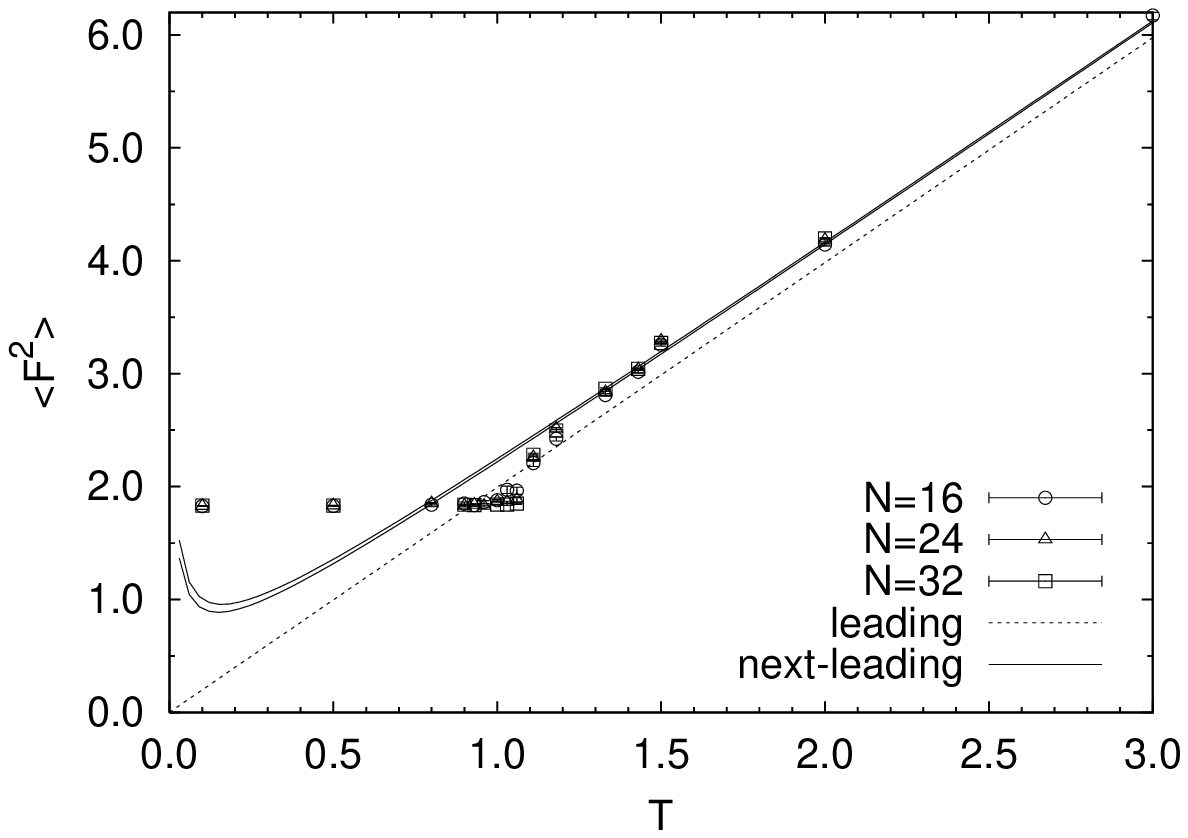, width=7.4cm}
\caption{
The observables $\langle R^2 \rangle$ and 
$\langle F^2 \rangle$ are plotted against $T$
% $(a=0.05,\,N_t=1/(0.05T))$  
for $\alpha=0.0$ and $N=16,24,32$\,.
The dashed lines represent the results obtained
by eqs.\ (\ref{R2DR}) and (\ref{F2DR}) in the high
$T$ limit.
%for the dimensionally reduced model (\ref{DRmodel}). 
The solid lines represent the results
including the next-leading order terms \cite{highT}.
}
\label{Fig_N16_alp_0o0}}

In figure \ref{Fig_N16_alp_0o0} we plot 
the observables $\langle R^2 \rangle$ and 
$\langle F^2 \rangle$ against $T$ at $\alpha =0$
for $N=16,24,32$.
The results for different $N$ lie on top of each other,
which implies a clear large-$N$ scaling behavior.
In the confined phase $T < T_{\rm H}$, we find that
the results
%the observables (\ref{DefFMR}) 
are independent of $T$.
This can be considered as a consequence of 
the Eguchi-Kawai equivalence,\footnote{\label{fuzzy-EK}Let us note 
that the results in the fuzzy sphere phase are
also consistent with Eguchi-Kawai's statement.
If we fix $T$ rather than $\tT$ in the large-$N$ limit,
the Polyakov line vanishes identically, and all the observables have
no dependence on $T$.
On the other hand, if we fix $\tT$ in the large-$N$ limit,
%as we did in the previous section, 
the Polyakov line vanishes only at $\tT=0$, 
and all the observables have non-trivial dependence on $\tT$.
}
which states the volume independence of single-trace
operators in $D$-dimensional
U($\infty$) gauge theory provided that
the U(1)$^D$ symmetry is not spontaneously broken \cite{EK}.

We have performed a similar analysis at
$\tilde{\alpha}=1.8$, which is barely below the 
boundary of the fuzzy sphere phase.
(See figure \ref{fig:phase-alpha-T}.)
The Hagedorn temperature turned out to be
$T_{\rm H} \sim 1.1$ as well.
In the Yang-Mills phase,
the Chern-Simons term $M$ takes small values
as one can see 
from figure \ref{fig:Phase_Trans_lower}, 
and the observables have little dependence on $\alpha$.
This property is found also
in the totally reduced model studied in ref.\ \cite{0401038}.
Note also that the Hagedorn temperature $T_{\rm H}$ is an O(1)
quantity, which means that 
$\tilde{T}_{\rm H} \equiv N^{-2/3} T_{\rm H}$
vanishes in the $N \rightarrow \infty$ limit.
In other words, if we drew the critical line
corresponding to the Hagedorn transition
in figure \ref{fig:phase-alpha-T},
it would be pushed towards the $\tilde{T}=0$ line
in the large-$N$ limit.
This is simply a reflection of the fact that,
in the fuzzy sphere phase, 
we have to consider super high temperature
to see non-trivial temperature dependence.
%the thermal effects.

%% \FIGURE{
%%  	    \epsfig{file=N32_alp1o8_PL.eps, width=7.4cm}
%%         \epsfig{file=N32_alp1o8_X2.eps, width=7.4cm}
%% \caption{$\langle P^2 \rangle$ and $\langle R^2 \rangle$ 
%% are plotted against  
%% $T\,(a=0.05,\,N_t=1/(0.05T))$  
%% for $\alpha \simeq 0.71\,(\tilde{\alpha}=1.8)$ and $N=32$\,.}
%% \label{Fig_N=16_alp=1o8}}

%% From these results, it is concluded that 
%% the Hagedorn temperature $T_H$ in the region $\alpha=0.71$ as  
%% \begin{align}
%% T_H = 1.0\pm 0.01\,.
%% \label{T_H at apper point}
%% \end{align}

\section{Fuzzy-sphere/Yang-Mills transition at high temperature}
\label{sec:dim-red}

In general, field theories at high temperature are effectively
described by bosonic field theories in one dimension less.
This phenomenon provides a useful approach to
QCD at high temperature.
(See, for instance, refs.\ 
\cite{Reisz:1991er,Bialas:2004gx} and references therein.)
In the present model,\footnote{The 
high temperature limit
in matrix quantum mechanics is also discussed 
refs.\ \cite{Bal-Sathiapalan,Aharony:2004ig,highT}.}
we do not have the subtlety related to infrared 
divergences unlike in ordinary field theories, 
since there is no infinitely extended spatial directions 
from the outset.
%% Thus 
%% %in the high temperature limit of the present model,
%% we obtain a totally reduced model,
%% which has been a subject of intensive studies.
%% We clarify the relationship to the results obtained in 
%% the previous works.

%Demonstration by Monte Carlo simulation}

The dimensionally reduced model is obtained from the 
original action (\ref{action}) by suppressing the $t$ 
dependence of the 1d fields as
\beqa
S_{\rm DR} &=&
\frac{N}{T}
\Bigl\{ -\frac{1}{2}([A, X_i])^2
-\frac{1}{4}\Bigl([X_i,X_j]\Bigr)^2
+\frac{2}{3} \, i \, \alpha \, 
 \varepsilon_{ijk} X_i X_j X_k \Bigr\} \nn \\
&=& N
%\frac{N}{T}
\Bigl\{ -\frac{1}{4}\Bigl([A_\mu,A_\nu]\Bigr)^2
+\frac{2}{3} \, i \, \gamma \, 
\varepsilon_{ijk}A_i A_j A_k \Bigr\} \ ,
\label{DRmodel}
\eeqa
where the Greek indices $\mu$, $\nu$ run over 
$1, \cdots ,4$ and we have defined
%$A_i \equiv T^{-1/4}X_i$ ($i=1,2,3$), $A_4 \equiv T^{-1/4}A$
\beqa
A_i & = & T^{-1/4}X_i \quad (i=1,2,3) \ , 
\quad A_4 = T^{-1/4}A  \ ,\\
\gamma &=& T ^{-1/4} \alpha \ .
\label{gamma-alpha}
\eeqa
%For instance, 
%the high temperature behavior of
The observables studied in the previous sections can be obtained 
at high temperature as
\beqa
\label{R2DR}
\langle R^2 \rangle &\simeq& 
 T^{1/2} \cdot \left\langle \frac{1}{N}\,
\tr (A_i)^2 \right\rangle_{{\rm DR},\gamma}  \ ,
\\
\label{M-DR}
\langle M \rangle &\simeq& 
 T^{3/4} \cdot
\left\langle 
\frac{2 \, i}{3 \, N} \, 
\ijk \, {\rm tr}  (A_i A_j A_k)
 \right\rangle_{{\rm DR},\gamma}  \ , \\
\langle F^2 \rangle &\simeq& 
-  T \cdot \left\langle 
\tr \Bigl( [A_i, A_j] \Bigr)^2 \right\rangle_{{\rm DR},\gamma}
\label{F2DR}  \ , \\
\langle |P| \rangle &\simeq&  
1- \frac{1}{2} \, T^{-3/2} \cdot
\left\langle 
\frac{1}{N}\, \tr
 (A_4)^2
\right\rangle_{{\rm DR},\gamma}   \ .
\label{polDR}
\eeqa
%$C=2.162(5)$, and 
The symbol
$\langle \ \cdot \ \rangle_{{\rm DR},\gamma}$ 
represents the expectation value with respect
to the dimensionally reduced model (\ref{DRmodel}),
where $\gamma$ is related to $\alpha$ through (\ref{gamma-alpha}).

%In the Yang-Mills phase, on the other hand, 
%the dimension reduction is complete as long as $T \gg 1$.
In the $\alpha=0$ case,
the corresponding dimensionally reduced model (\ref{DRmodel})
is studied in detail at large $N$ \cite{HNT}.
For instance, we have
\beqa
 C \equiv
\lim_{N \rightarrow \infty}
\left\langle \frac{1}{N}\tr (A_\mu)^2 \right\rangle_{{\rm DR},0}
&=&  2.162(5)  \ , \\
- \left\langle \frac{1}{N}
\tr \Bigl( [A_\mu, A_\nu] \Bigr)^2 \right\rangle_{{\rm DR},0}
&=& 4  \left( 1-\frac{1}{N^2} \right)  \ . 
\eeqa
Taking into account that the Greek indices
run from 1 to 4 in contrast to the Roman indices,
which run from 1 to 3,
we obtain the asymptotic behavior of 
the original model with $\alpha = 0$ at high $T$ as
\beqa
\lim_{N\rightarrow \infty}
\langle R^2 \rangle &\simeq& \frac{3}{4} C \sqrt{T} \ , \\
\langle F^2 \rangle &\simeq& 2 \, T \left( 1-\frac{1}{N^2} \right) \ . 
\eeqa
Figures
\ref{Fig_N16_alp_0o0_PL} and \ref{Fig_N16_alp_0o0}
%\ref{Fig_N16s_alp_0}
show that our Monte Carlo results approach
%tend to agree with
these results at high $T$.
(Small deviations can be nicely reproduced by the 
next-leading order calculation \cite{highT}.)

%%%%%%%%%%%%

In the fuzzy sphere phase,
we can confirm the dimensional reduction analytically
by using the all order calculation in perturbation theory.
By taking the $\tb \rightarrow 0 $ limit in
the results for the full model, we obtain 
the all order results for the dimensionally reduced model,
which can be obtained similarly to ref.\ \cite{ANN}.
In contrast to the situation in the Yang-Mills phase,
$\tT$ instead of $T$ has to be large (in the large-$N$ limit)
in order for the dimensional reduction to take place.
%in the fuzzy sphere phase.

%\section{The order of phase transition}
%\label{sec:order-transition}

\FIGURE{
%\epsfig{file=PT_CS_1.eps,width=7.4cm}
% results for the CS term :   CS_upper_T30o0_2.eps
\epsfig{file=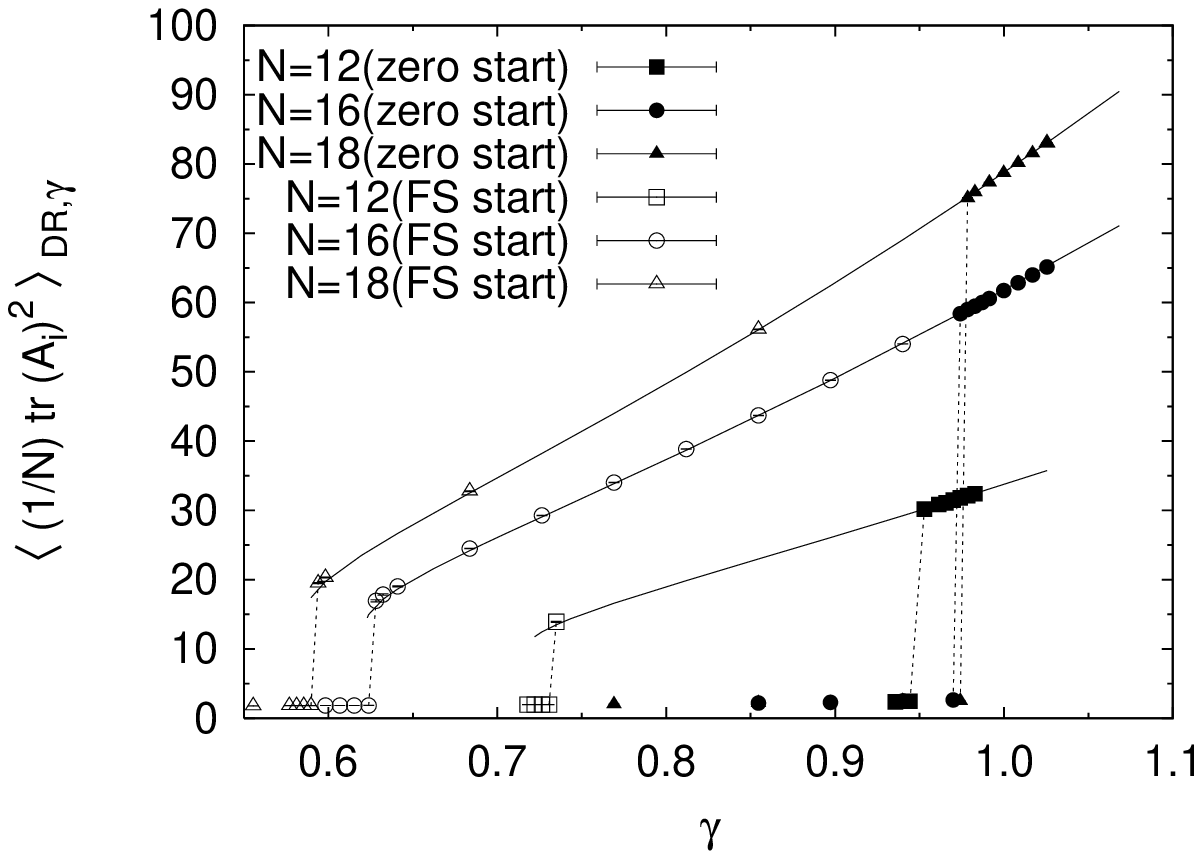,width=7.4cm}
\epsfig{file=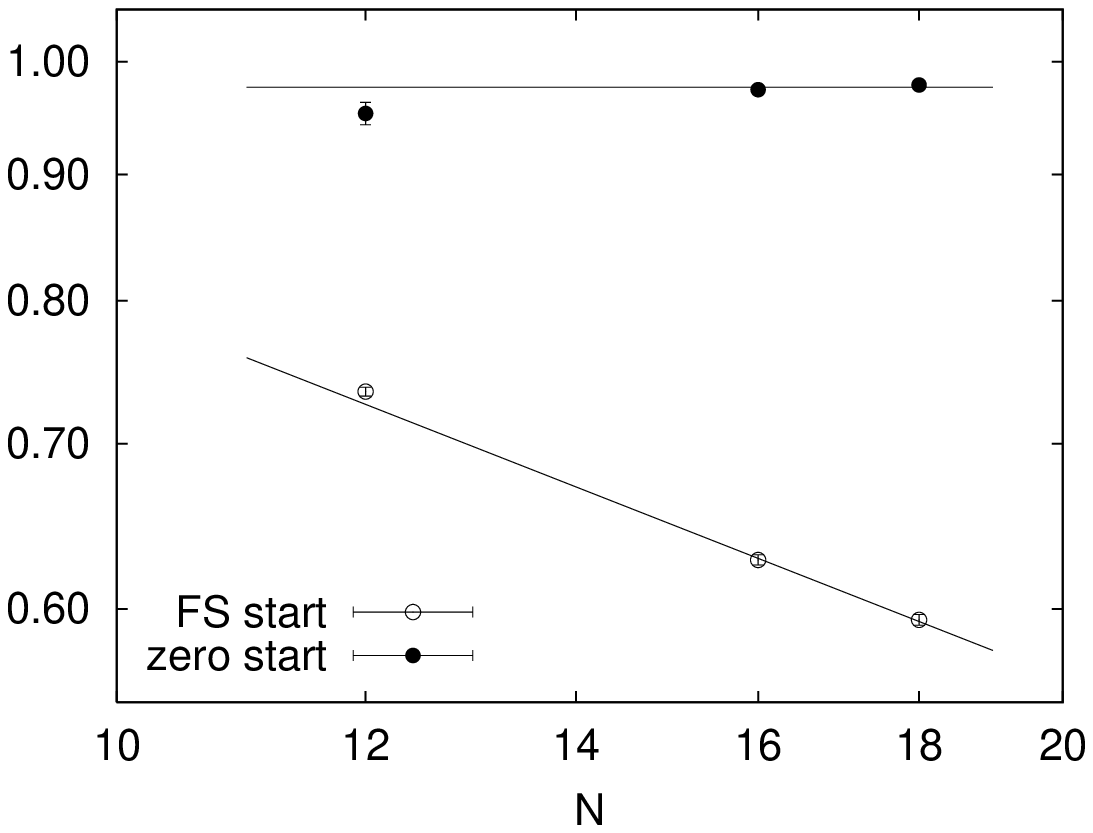, width=7.4cm}
\caption
{(Left) The observable
$\left\langle \frac{1}{N}\,
\tr (A_i)^2 \right\rangle_{{\rm DR},\gamma}$
%% and 
%% $\left\langle 
%% \frac{2 \, i}{3 \, N} \, 
%% \ijk \, {\rm tr}  (A_i A_j A_k)
%%  \right\rangle_{{\rm DR},\gamma}$
in the dimensionally reduced model
is plotted against $\gamma$ for $N=12,16,18$.
%         $\langle M \rangle / N$ against $\tilde{\alpha}$ and $N=16$ 
%         for $T=30.0$\,.
The open and closed symbols represent the results for
the single fuzzy sphere start
and the zero start, respectively.
%         The closed represent the zero start.
%         The doted line represents the classical result. 
%         The dashed line represents the one-loop result.
         The solid lines represent the all order results.
(Right) The upper and lower critical points represented
by closed and open circles, respectively, are plotted
against $N$ in the log-log scale.
The straight lines represent the fits to
$\gamma^{\rm (u)}_{\rm cr} = c_1$ and 
$\gamma^{\rm (l)}_{\rm cr} = c_2 N^{-1/2}$,
where $c_1 = 0.9765$ and $c_2=2.5160$.
}
\label{Phase_Trans_upper}}

Using the dimensionally reduced model
(\ref{DRmodel}), let us investigate
the phase transition between the
fuzzy sphere phase and the Yang-Mills phase
in the high temperature limit.
This clarifies, in particular, the first
order nature of the phase transition,
and it also enables us to make explicit the connection
to the known results in a totally reduced model \cite{0401038}.
We perform Monte Carlo simulation\footnote{We have used
the same algorithm as in ref.\ \cite{0401038}.}
of the dimensionally
reduced model (\ref{DRmodel})
% to obtain results for the full model at high temperature.
%
%% As an application of the dimensional reduction,
%% let us study
%% the order of the phase transition between the
%% fuzzy sphere phase and the Yang-Mills phase.
%% Although Monte Carlo simulation of the full model
%% can be used in principle for that purpose, it
%% is actually technically difficult
%% since the present model possesses
%% many local minima of the effective action
%% corresponding to the classical solutions 
%% discussed in section \ref{sec:model}.
%% However, we found that the thermalization time 
%% is considerably reduced in the dimensionally reduced model.
%% Therefore, we address the issue
%% in the high temperature regime, where the 
%% dimensional reduction is justified.
%% %by simulating the dimensionally reduced model.
using, as the initial configuration,
either of the two configurations given by
\beq
A_i=
\left\{
\begin{array}{ll}
\gamma \, L_i^{(N)} \quad& (\textrm{the single fuzzy sphere start})\ ,\\ 
0
%{\bf 1}_N 
\quad & (\textrm{the zero start})\ ,
\end{array}
\right.
\eeq
and $A_4=0$ for both cases.
In figure \ref{Phase_Trans_upper} (Left)
we plot the observable appearing on the right hand side of
eq.\ (\ref{R2DR})
against $\gamma$ for $N=12,16,18$.
For comparison
we also plot
the all order results
for the dimensionally reduced model
obtained from the perturbation theory 
around the single fuzzy sphere in the large-$N$ limit.
The Monte Carlo results depend on the initial configuration in the
intermediate region of $\gamma$, and we observe
discontinuities at
\begin{equation}
\gamma = \left\{
\begin{array}{rcll}
\gamma_{\rm cr}^{\rm (l)} &\sim& \frac{2.5}{\sqrt{N}}\qquad & 
{\mbox{for the single fuzzy sphere start}}  \ ,
\\
\label{criticalpointzero}
\gamma_{\rm cr}^{\rm (u)} &\sim& 0.98
& {\mbox{for the zero start}} \ ,
\end{array} 
\right.
\end{equation}
which we call the lower/upper critical points, respectively.
(See figure \ref{Phase_Trans_upper} (Right) for a plot showing
the large-$N$ behaviors.)  
This clearly demonstrates that the phase transition
is of first order.
%The lower critical point $\gamma_{\rm cr}^{\rm (l)}$
%obtained by Monte Carlo simulation agrees well with the
%result obtained from the one-loop effective action
%for the dimensionally reduced model.

In a similar model \cite{0401038}, which can be
obtained by simply omitting $A_4$ from (\ref{DRmodel}),
the critical points are obtained as
$\gamma_{\rm cr}^{\rm (l)} \sim \frac{2.1}{\sqrt{N}}$
and $\gamma_{\rm cr}^{\rm (u)} \sim 0.66$, respectively.
We find that the inclusion of the fourth matrix $A_4$
changes the numerical coefficients, but not the powers of $N$,
in the large-$N$ behavior of the critical points.

%There are two transition points depending on the initial
%configuration, and in between 
%we observe a clear hysteresis behavior.
%We observe a clear hysteresis behavior in the intermediate region
%of $\alpha$, 

Using the relation (\ref{gamma-alpha}),
we obtain the critical points in terms of the parameters
of the full model as
\begin{equation}
\alpha = \left\{
\begin{array}{rcl}
\alpha_{\rm c}^{\rm (l)} &\sim& \frac{2.5}{\sqrt{N}} 
\, T^{1/4} \ , \qquad 
\\
\label{criticalpointzero2}
\alpha_{\rm c}^{\rm (u)} &\sim& 0.98 \, T^{1/4} \ .
\end{array} 
\right.
\end{equation}
In this terminology, the critical point $\ta_{\rm c}$
shown in figure \ref{fig:phase-alpha-T} is actually the
{\em lower} critical point.
Note that the factor $\frac{1}{\sqrt{N}}$ in
(\ref{criticalpointzero2}) is absorbed by the
rescaling (\ref{rescalded-var}) of $\alpha$ and $T$, and 
the result agrees with the high $\tT$ behavior (\ref{ta_highT}) 
obtained from the effective action.

%%%%%%%%%%%%%%%%%%%%%%%%%%%%%%%%%

\section{Summary and discussions}
\label{sec:summary}

We have studied thermodynamical properties
of a fuzzy sphere in
a BFSS-type matrix model including
the Chern-Simons term.
We have established the phase diagram
in the $(\alpha , T)$-plane,
and obtained, in particular,
the phase boundary between the fuzzy sphere
phase and the Yang-Mills phase
as shown in figure \ref{fig:phase-alpha-T}.

In the fuzzy sphere phase, we are able to obtain
all order results for various observables
exploiting the one-loop saturation of the effective 
action in the large-$N$ limit.
This technique was previously applied to
various fuzzy manifolds in totally reduced models.
We consider it interesting
that it can be generalized to a finite temperature
setup in a straightforward manner.
Following refs.\ \cite{0405277,0506205}
thermodynamical properties of four-dimensional
fuzzy manifolds such as fuzzy ${\rm CP}^2$ and
fuzzy ${\rm S}^2 \times {\rm S}^2$
can be studied in a similar way.

%% We have also performed Monte Carlo simultion.
%% In general results obtained from simulations contain
%% both statistical error and systematic error.
%% The sources of the latter error comes from the finite
%% lattice spacing used for discretization of the Euclidean time.
%% We have taken the lattice spacing sufficiently small 

One of the interesting aspects of our results is
the scaling of parameters in the large-$N$ limit.
In the fuzzy sphere phase, if one fixes the original parameters
$\alpha$ and $T$ in the large-$N$ limit, one simply obtains
trivial results corresponding to 
the classical fuzzy sphere at zero temperature.
In order to keep non-trivial quantum corrections and 
thermal effects,
one has to hold $\ta$ and $\tT$ fixed in the large-$N$ limit.

In that limit, we find that
the Polyakov line vanishes smoothly as $\tT$ approaches $0$.
This implies that the fuzzy sphere phase is not
further divided into the confined phase and the deconfined phase,
unlike the Yang-Mills phase.
If we take the large-$N$ limit at fixed $T$, we are
always in the confined phase.
If we take the large-$N$ limit at fixed nonzero $\tT$, 
we are always in the deconfined phase.
In ref.\ \cite{FSS} it is stated that 
the Hagedorn temperature for the fuzzy sphere
is $T_{\rm H} = \infty$ in an analogous model.
We consider that our results provide a more 
precise formulation of that statement.

%% We have also demonstrated that the dimensional reduction
%% takes place at sufficiently high temperature.
%% In the fuzzy sphere phase, it occurs for sufficiently large $\tT$,
%% whereas in the Yang-Mills phase, it occurs for sufficiently large $T$.

As an outlook, we note that fuzzy manifolds 
\cite{0207115,0303120,0506033}
are also studied intensively in the
IIB matrix model \cite{9612115}
in order to investigate the dynamical generation of 4d space-time.
The same issue has been addressed by 
various approaches \cite{Aoki:1998vn,Ambjorn:2000bf,Ambjorn:2000dx,%
NV,exact,sign,Nishimura-Sugino,SSB},
and in ref.\ \cite{Nishimura-Sugino} the first evidence for such
a phenomenon is obtained by the Gaussian expansion method.
Based on the Eguchi-Kawai equivalence \cite{EK},
two of the authors (N.K.\ and J.N.) 
conjectured \cite{Kawahara:2005an}
that a similar phenomenon should occur
in the BFSS matrix model \cite{9610043}.
We therefore consider that studying the effective action for
fuzzy manifolds in the BFSS matrix model 
would be an interesting future direction.
In that case, the effective action is expected to be
saturated at two loop similarly to the situation in 
the IIB matrix model \cite{0403242}.

From the view point of the gauge/gravity correspondence,
the fuzzy sphere solutions 
in the pp-wave matrix model
can be interpreted as giant gravitons.
It would be interesting to look for 
phenomena in the dual gravity theory 
corresponding to the ones discussed in this paper.

%\vspace*{0.5cm}
\section*{Acknowledgments} 

We would like to thank
Takehiro Azuma, Kazuyuki Furuuchi,
Yoshihisa Kitazawa,
Shun'ya Mizoguchi, Kentaroh Yoshida
and Gordon Semenoff for valuable comments and 
discussions.
The simulations were performed on the PC clusters 
%with 25 nodes of Pentium4 (2.)
at KEK.

\end{document}